\documentclass[lettersize,journal]{IEEEtran}
\usepackage{amsmath,amsfonts}
\usepackage{algorithmic}
\usepackage{array}
\usepackage[caption=false,font=normalsize,labelfont=sf,textfont=sf]{subfig}
\usepackage{textcomp}
\usepackage{stfloats}
\usepackage{url}
\usepackage{verbatim}
\usepackage{graphicx}
\usepackage{cite}
\usepackage{float}  
\usepackage[linesnumbered]{algorithm2e} 
\usepackage{threeparttable} 
\usepackage{multirow}

\begin{document}

\title{Pro-Prophet: A Systematic Load Balancing Method for Efficient Parallel Training of Large-scale MoE Models}

\author{\IEEEauthorblockN{Wei Wang, Zhiquan Lai, Shengwei Li, Weijie Liu, Keshi Ge, Ao Shen, Huayou Su, Dongsheng Li} \\
\thanks
{The first two authors contributed equally to this work.}
\thanks{Wei~Wang, Zhiquan~Lai, Shengwei~Li, Weijie~Liu, Keshi~Ge, Ao~Shen, Huayou~Su and Dongsheng~Li are with the National Key Laboratory of Parallel and Distributed Computing, College of Computer, National University of Defense Technology in Changsha, Hunan, China. (Corresponding author: Dongsheng Li) \\
E-mail: \{wwking, zqlai, swli, liuweijie, gekeshi, shenao, shyou, dsli\}@nudt.edu.cn
}
\thanks
{This work is supported by the National Key R\&D Program of China (No. 2022YFB4501400) and the National Natural Science Foundation of China under Grant No. 62025208 and 62421002.}
}

\maketitle

\begin{abstract}
The size of deep learning models has been increasing to enhance model quality. The linear increase in training computation budget with model size means that training an extremely large-scale model is exceedingly time-consuming. Recently, the Mixture of Expert (MoE) has drawn significant attention as it can scale models to extra-large sizes with a stable computation budget. However, inefficient distributed training of large-scale MoE models hinders their broader application.
Specifically, a considerable dynamic load imbalance occurs among devices during training, significantly reducing throughput. Several load-balancing works have been proposed to address the challenge. System-level solutions draw more attention for their hardware affinity and non-disruption of model convergence compared to algorithm-level ones. However, they are troubled by high communication costs and poor communication-computation overlapping.
To address these challenges, we propose a systematic load-balancing method, Pro-Prophet, which consists of a planner and a scheduler for efficient parallel training of large-scale MoE models. 
To adapt to the dynamic load imbalance, we profile training statistics and use them to design Pro-Prophet.
For lower communication volume, Pro-Prophet planner determines a series of lightweight load-balancing strategies and efficiently searches for a communication-efficient one for training based on the statistics.
For sufficient overlapping of communication and computation, Pro-Prophet scheduler schedules the data-dependent operations based on the statistics and operation features, further improving the training throughput.
We conduct extensive experiments in four clusters and five MoE models. The results indicate that Pro-Prophet achieves up to 2.66x speedup compared to two popular MoE frameworks including Deepspeed-MoE and FasterMoE. Furthermore, Pro-Prophet has demonstrated a load-balancing improvement of up to 11.01x compared to a representative load-balancing work, FasterMoE.
\end{abstract}


\section{Introduction}
\IEEEPARstart{R}{ecent} years, large-scale deep neural networks have achieved superior performance in various domains(e.g., NLP, CV). 
Previous works have shown that the model capacity is improved with the increased model size, further promoting the model scaling. However, the substantial computational demand of extra-large models makes the training process excessively time-consuming. As one of the most promising solutions, Mixture of Expert (MoE) enables a nearly constant computational budget as model scaling. 
Generally, we replace some layers of a foundation model with MoE ones to generate a MoE model. Each MoE layer contains a gate network and a range of sub-modules named \textit{experts}. The gate network can route each input to top-$k$ experts that excel in processing the input. As the $k$ is a super-parameter, the MoE model can be scaled with consistent computational requirements by increasing the number of experts.

As the model further scales, the effective collaboration of devices is necessary for extra-large MoE model training. Unfortunately, it is inefficient to train the model with traditional parallelism such as Data Parallelism (DP), Model Parallelism (MP), and Pipeline Parallelism (PP). To overcome the trouble, Gshard~\cite{gshard} introduced a specific parallel strategy named \textit{Expert Parallelism} (EP). Nowadays, extra-large MoE models trained with EP have demonstrated the highest accuracy in multiple tasks\cite{glam, liu2024deepseek}.

However, training MoE models using EP presents a dynamic load imbalance among devices. For each MoE layer, EP equally divides experts into devices before training and dynamically arranges its inputs according to the gate network during training. Most inputs are transferred to and processed by a few devices, resulting in prolonged communication and computation of inputs. Furthermore, the imbalance varies throughout the training process, making it difficult to resolve. 

Numerous attempts in load-balancing have been proposed to improve training throughput. 
Algorithmic works often restrict the upper bound of each expert's load~\cite{deepspeed, tutel} or add auxiliary losses to the loss function~\cite{switch, 2017MoE} for a more balanced load. However, they impact the model convergence and even deteriorate the model quality.
Considering the drawback above, systematic solutions of the MoE system draw more attention. Popular systematic works~\cite{fastermoe, nie2023flexmoe} dynamically readjust the expert placement according to the load, achieving a balanced load without harming the model quality.


However, these systematic solutions struggle to enhance training efficiency effectively due to two drawbacks.
1) Heavy communications of model states (i.e., parameters, gradients and optimizer states~\cite{rajbhandari2020zero}) are introduced. 
The previous expert placements introduce a global transfer of parameters and gradients or a whole model states communication.
These transferring strategies involve unnecessary communications across devices, hindering the improvement of training efficiency.
2) Devices experience significant communication and computation idle during training. 
Due to data dependencies among operators, the solutions have to perform some communications and computations sequentially.
For example, only the experts have been selected and their model states have been transmitted, their computations of inputs can be launched. And then, the aggregation of gradients occurs only after the computation of gradients is finished. 
They neglect the potential of communication and computation overlapping, thus significantly influencing device utilization.

In this paper, we propose a systematic load-balancing solution, Pro-Prophet, which overcomes two drawbacks by a planner and scheduler respectively. 

To adapt dynamic features presented in the training of a MoE model, we profile the \textit{input distribution} (i.e., the number of inputs processed by each expert) for each MoE layer. We observe that distributions of a MoE layer between adjacent iterations present high similarity. This \textit{locality} is the key to effective load balancing.

To reduce the communication volume, Pro-Prophet planner introduces a series of \textit{lightweight expert placements}. In a lightweight expert placement, each expert is independently allocated to a subset of devices. Communication of parameters and gradients for the expert occurs in these specific devices. Serve to evaluate expert placements, the planner proposes a \textit{performance model} that estimates the execution time of a MoE layer employing a lightweight expert placement. However, it is non-trivial to find the optimal one due to the combinatorial explosion of the number of expert placements. To tackle this, the planner designs a \textit{locality-based greedy algorithm}. The algorithm employs a greedy strategy to search for a communication-efficient expert placement. Besides, its launching frequency is reduced based on the locality, further improving the training throughput.  


To exploit the potential of communication-computation overlapping, Pro-Prophet scheduler comprehensively schedules operations based on the locality and the feature of operations.
The locality means that we can estimate the input distribution of the upcoming iteration according to the current one. Once the upcoming distribution is obtained, we can promptly determine a communication-efficient expert placement for the upcoming iteration and can transmit the parameters of experts in advance, which provides the opportunity to overlap communications and computations within adjacent iterations. Besides, the gradient aggregation can be scheduled backward for better overlapping. 
Based on these, the scheduler identifies a scheduling space and designs a \textit{block-wise scheduling strategy} to comprehensively overlap communications and computations.

We implement Pro-Prophet on top of PyTorch and conduct extensive experiments on four different clusters of up to 32 devices with five variant models.
The results demonstrate that Pro-Prophet achieves speedups of up to 2.66x compared to two popular MoE frameworks. Additionally, Pro-Prophet has demonstrated load-balancing enhancements of up to 11.01x compared to a representative load-balancing work, FasterMoE.

Our main contributions are summarized as follows:
\begin{enumerate}
\item[$\bullet$] We profile input distributions among adjacent iterations and identify a locality that guides the design of Pro-Prophet.
\item[$\bullet$] We design a Pro-Prophet planner that identifies several lightweight expert placements, abstracts a performance model and designs a locality-based greedy algorithm to reduce the heavy communication of model states.
\item[$\bullet$] We propose a Pro-Prophet scheduler, which generates a scheduling space and establishes a block-wise scheduling strategy based on the locality and the feature of operations for comprehensive overlapping of computations and communications.
\item[$\bullet$] We conduct comprehensive experiments for Pro-Prophet on different clusters and models. The results demonstrate that Pro-Prophet achieved up to 1.50x end-to-end speedup and 11.01x load-balancing enhancements with the representative load-balancing method.
\end{enumerate}

\section{Background and Motivation}\label{sec:back_moti}

\subsection{Background}
Recent works in DNN model training have shown that the model capacity can be improved with increasing training data, model scale, and computational budget~\cite{kaplan2020scaling}. Extraordinary performance has been achieved in several deep learning domains including natural language processing (NLP), computer vision (CV), and so on.

However, significant training overhead comes along with the superior model capacity. The extra large-scale model~\cite{bert, T5, XLNet, roberta, GPT2, GPT3} training often takes months on thousands of dedicated accelerators (e.g., MT-NLG\cite{megatron-nlg} spends three months to train on over two thousand A100 GPUs), which influences the development of deep learning.

In recent years, dynamic sparse-activated architectures have been proposed to solve the trouble. One of the popular approaches is the Mixture of Experts (MoE), which can significantly improve the model capacity while maintaining a consistent computational budget. Nowadays, MoE has been successfully applied to large language models\cite{he2024mixture, zuo2021taming, ludziejewskiscaling, komatsuzaki2022sparse, xue2022go, xue2022one, zoph2022st}. Excellent MoE models that appeared in industry and academia greatly draw researchers' attention. For example, Google has trained a series of MoE models called Glam\cite{glam}. The largest Glam model is seven times larger than GPT-3, but the training cost is less than 1/3 of it. Experiments show that these models achieve higher accuracy than GPT-3 in 29 zero, single, and small sample learning tasks, representing the superiority of MoE models.
The other example is GPT-4\cite{achiam2023gpt}. The technical report of OpenAI indicates that the GPT-4 is a MoE model, which achieved the highest performance in various downstream tasks. Besides, the ChatGPT based on the GPT-4 has caused a tremendous sensation.  

\begin{figure}[tbp]
\centerline{\includegraphics[width=1.0\linewidth]{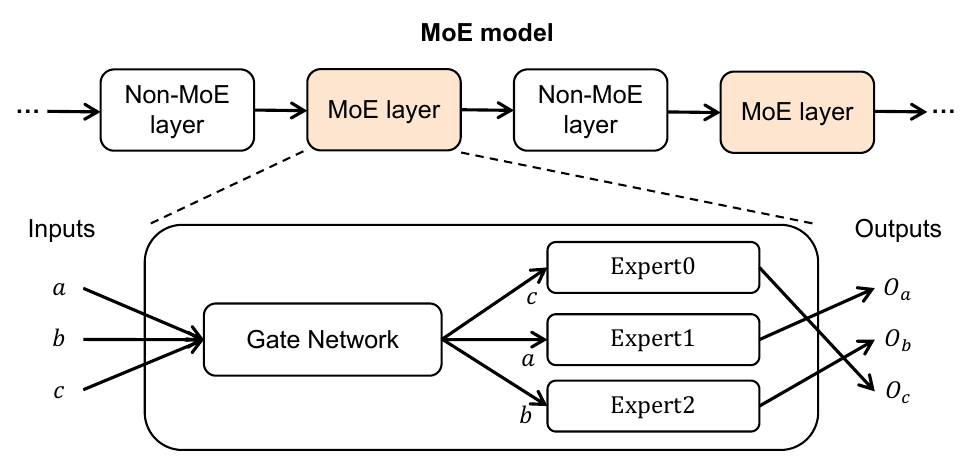}}
\caption{The structure of a MoE model and MoE layer. The MoE model consists of both MoE and non-MoE layers stacked on top of each other.
The MoE layer consists of a series of experts and a gate network for routing input to experts. For each input, the gate network computes the relationship between the input with three experts and allocates it to the top-$1$ expert for computation.}
\label{fig:MoE_structure}
\end{figure}

Fig.~\ref{fig:MoE_structure} illustrates the architecture of a MoE model and a MoE layer. The MoE model comprises a stack of non-MoE and MoE layers. A MoE layer consists of two components: 1)\ a series of experts (3 experts in the figure), where each excels in a specific domain. 
2)\ a gate network, which routes each input to a few experts that are skilled in dealing with this input, rather than all experts.
In a MoE layer, for each input, the gate network computes the relationship between that input and all the experts. Then it routes the input to top-$k$ ($k$=1 in the figure) expert(s) for computation. 
Even as we increase the number of experts (the model size is increased), each input is still routed to a fixed number ($k$) of experts, and the negligible increased computational budget of the gate network, thereby enabling the scaling of the model with nearly constant computational overhead.

With the increase of the model scale, an isolated device cannot support the training of the MoE model thus various parallelisms have been proposed. Two common parallel approaches are DP and MP. DP equally divides inputs of an iteration across all devices and replicates the model into all devices. In forward propagation (FP), each device computes its local inputs independently by utilizing its model replicas. In the backward propagation (BP), the Allreduce primitive will be performed after the backward computation.
Different from DP, MP partitions the model into devices in a specific manner and each device contains a complete copy of the data. The aggregation primitive will be launched whenever required in FP and BP.

\begin{figure}[tbp]
\centerline{\includegraphics[width=1.1\linewidth]{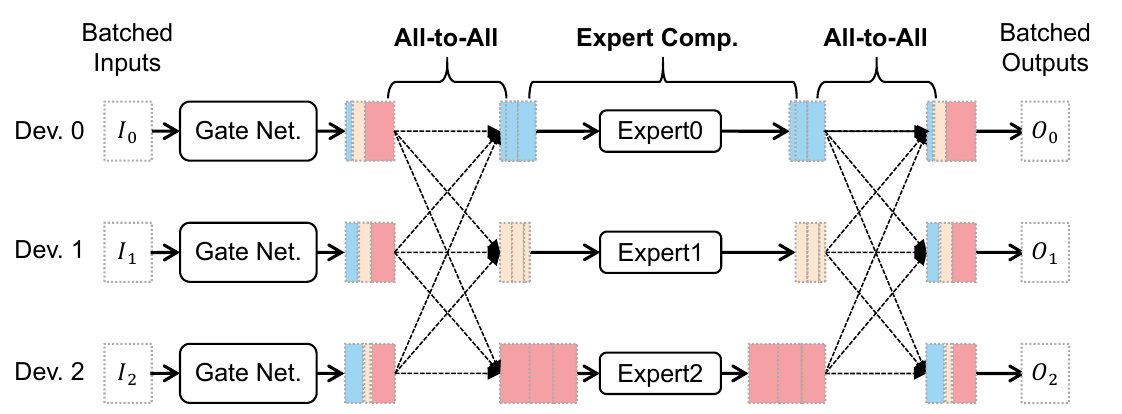}}
\caption{A workflow of Expert Parallelism (EP) in a MoE layer. Following the gate network, batched inputs are first exchanged via an All-to-All (A2A) operation. After the expert computation on all devices, a second A2A operation is used to pass the expert's outputs back to the device where corresponding inputs were originally located.}
\label{EP}
\end{figure}

For efficient training of a MoE model, Google combines DP and MP into an EP. From the input side, EP adopts the same input partitioning paradigm as DP. From the model side, EP divides the same number of experts to each device and copies other parts of the model (i.e., the gate network and non-MoE layer) to all devices.

Fig.~\ref{EP} illustrates a workflow of EP in a MoE layer.  
Firstly, the gate network determines top-$1$ expert for each input. Then inputs are transferred to corresponding devices via an All-to-All (A2A) communication operation\cite{A2A1, A2A2, A2A4}. Subsequently, each device performs the expert computation for collected inputs and then launches another A2A to reorganize the results back to the inputs' original devices for computations of the subsequent non-MoE layer. 
Nowadays, many popular distributed frameworks support the training of large-scale MoE models using EP\cite{megatron, deepspeed, nie2022hetumoe, he2021fastmoe, yu2024moesys}.

\begin{figure}[tbp]
\centerline{\includegraphics[scale=0.53]{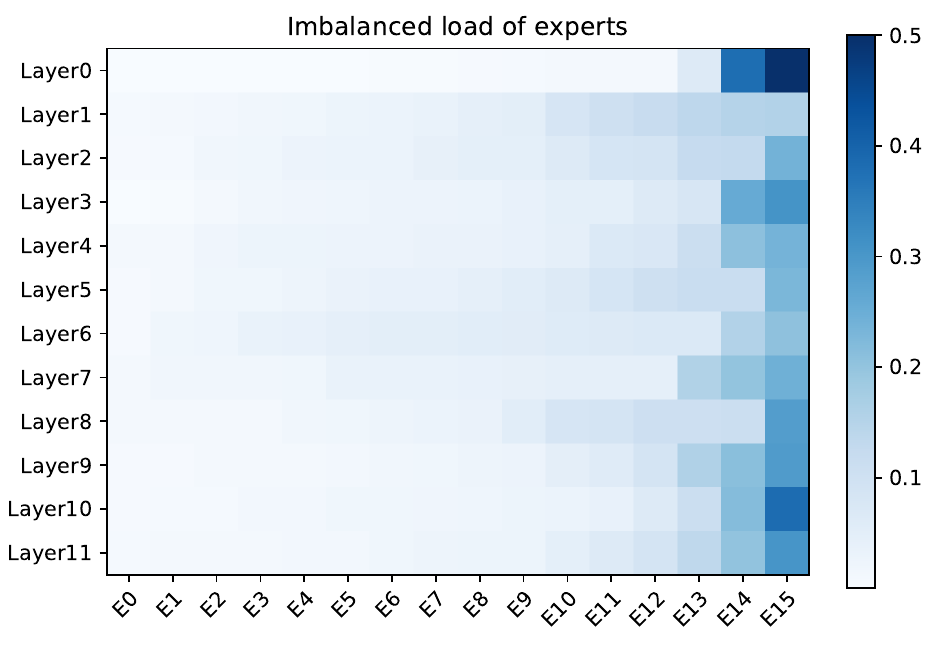}}
\caption{The imbalanced load of experts in an iteration. The model contains 12 MoE layers and each MoE layer contains 16 experts. The vertical axis indicates layer indexes, and the horizontal axis denotes the index of experts. The depth of color represents the proportion of total inputs that an expert handles. Three of the heaviest experts are responsible for over 50\% inputs while the three least experts only compute less than 5\%.}
\label{fig:load_imbalance}
\end{figure}

Even though EP makes it feasible to train extra-large MoE models with up to trillions of parameters, a dynamic load imbalance occurs among devices. 
Specifically, most of the training inputs are transferred and processed by a few devices. These heavy-load devices vary as the training.
Fig. \ref{fig:load_imbalance} presents the imbalanced load of experts in an iteration. The vertical axis indicates layer indexes, and the horizontal axis denotes the index of experts. The MoE model contains 12 MoE layers and each MoE layer contains 16 experts. Each expert is set into a dedicated device. The depth of color represents the proportion of total inputs processed by an expert.
In most MoE layers, the three heaviest experts hold over 50\% of inputs, while the three least less than 5\%. The unbalanced load of experts means that devices containing light-load experts have to wait for devices containing heavy-load ones, incurring significant under-utilization of devices during training.


\subsection{Motivation}

A series of methods have been proposed to balance the load. We divide them into algorithmic and systematic methods. From the algorithmic side, researchers constrain the upper bound of inputs received by an expert or add auxiliary losses to the loss function. They change the inputs-to-experts mapping, thus affecting and even deteriorating the model convergence. 



Different from algorithmic works, systematic solutions do not affect model convergence and fit into the hardware, thus attracting extensive attention. These solutions adaptively adjust experts-to-devices mapping based on the device load during training, effectively improving the training efficiency. However, their heavy load-balancing overhead hinders the further improvement of the efficiency. 


\begin{table}[htb]
\caption{Time breakdown of training. L.B. is short for Load balancing}
\centering
\scalebox{1.0}{
    \begin{tabular}{|c|c|c|c|c|c|c|}
    \hline
        \multicolumn{2}{|c|}{Model} & L.B. & Search & Place & Reduce & Others \\
    \hline
        \multicolumn{2}{|c|}{MoE-GPT-S} & 29.9\% & 6.8\% & 11.6\% & 11.5\% & 70.1\% \\
        \multicolumn{2}{|c|}{MoE-GPT-M} & 29.2\% & 3.2\% & 12.5\% & 12.5\% & 70.8\% \\
        \multicolumn{2}{|c|}{MoE-GPT-L} & 34.5\% & 2.6\% & 14.2\% & 17.7\% & 64.5\% \\
    \hline
        \multicolumn{2}{|c|}{MoE-GPT-DS} & 33.8\% & 6.1\% & 13.8\% & 13.9\% & 66.2\% \\
        \multicolumn{2}{|c|}{MoE-GPT-DM} & 37.1\% & 6.1\% & 16.1\% & 14.9\% & 62.9\% \\
    \hline
    \end{tabular}
}
\label{table:profile}
\end{table}

As shown in Table~\ref{table:profile}, previous solutions introduce a \textit{Search}, \textit{Place} and \textit{Reduce} processes to balance the load. However, the overhead of load-balancing is up to 37.1\%. There are two reasons behind this huge cost. Firstly, they do heavy communication of model states. They have to transfer the parameters and gradients of heavy-load experts among all devices or transmit the whole model states of experts. Secondly, they cannot sufficiently overlap the communication and computations during the training of a MoE model. Due to the data dependency, they have to do communication and computation sequentially. 

\begin{figure}[tbp]
\centerline{\includegraphics[width=1\linewidth]{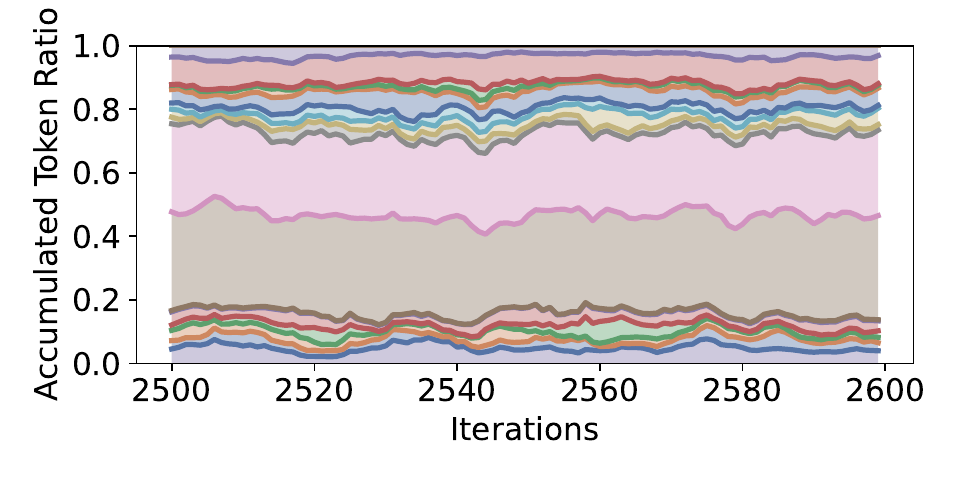}}
\caption{The locality of input distributions. The discrepancies between the different colored curves represent the number of inputs received by each of the different experts. It shows that distributions of adjacent iterations remain relatively constant.}
\label{fig:locality}
\end{figure}

\textbf{Locality.}
Fortunately, we have discovered a property in the training of MoE models. This property makes it possible to address these challenges efficiently.
Fig. \ref{fig:locality} depicts input distributions in the second MoE layer of a MoE model. The areas with different colors represent inputs received by different experts. It is worth noting that the distribution in other MoE layers follows a similar pattern. As shown in the figure, the slight fluctuation of the distribution occurs across adjacent iterations, which indicates that the load of each expert remains relatively stable in adjacent iterations. This phenomenon suggests that the distribution exhibits a locality among iterations.

\section{Overview of Pro-Prophet}\label{sec:overview}

\begin{figure}[tbp]
\centerline{\includegraphics[width=1.05\linewidth]{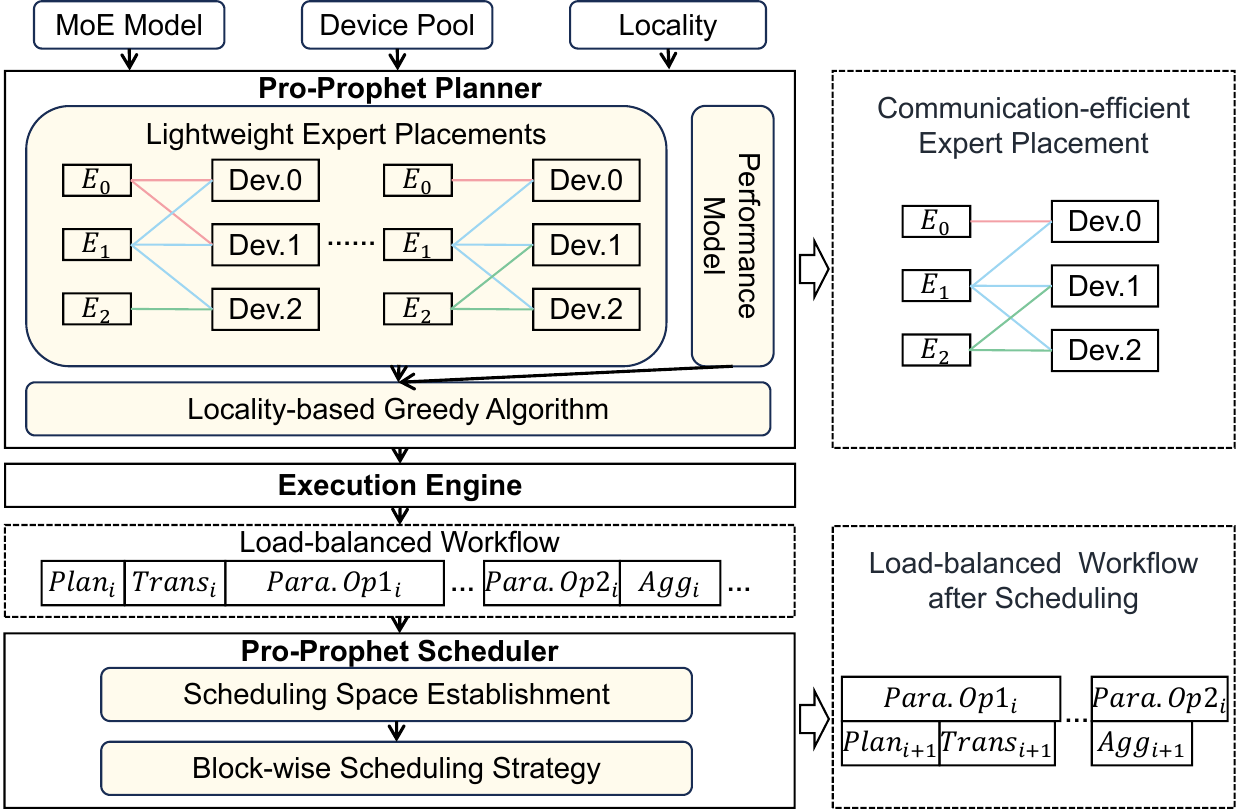}}

\caption{The overview of Pro-Prophet. Pro-Prophet is composed of Pro-Prophet planner and Pro-Prophet scheduler. MoE model, locality, and device pool are three inputs of it. Firstly, Pro-Prophet planner searches for a communication-efficient expert placement using its locality-based greedy algorithm. The algorithm iteratively generates and evaluates a lightweight expert placement utilizing its performance model until the load is balanced. Then the execution engine produces a load-balancing workflow based on the planner. Finally, Pro-Prophet scheduler schedules three data-dependent operations to parallel operations for communication and computation overlapping, further improving the training throughput.}
\label{fig:overview}
\end{figure}

Motivated by Section~\ref{sec:back_moti}, we propose a systematical load-balancing approach, Pro-Prophet, which can efficiently balance the load of devices. The overview of Pro-Prophet is presented in Fig.~\ref{fig:overview}. Pro-Prophet is composed of a planner and a scheduler. MoE model, locality, and device pool are three inputs of it. The device pool defines the topology of devices. The utilization of the locality is the key advantage of Pro-Prophet.

Firstly, Pro-Prophet planner searches for a communication-efficient expert placement from a series of lightweight expert placements using its locality-based greedy algorithm. The algorithm iteratively generates and evaluates a lightweight expert placement utilizing a performance model until the load is balanced. 

Then, the execution engine analyzes the procedures of the planner and produces a load-balanced workflow for load balancing. 

Finally, after analyzing the workflow, Pro-Prophet scheduler establishes the scheduling space and schedules data-dependent operations (i.e., \verb|Plan|, \verb|Trans|, and \verb|Agg|) to parallel operations (i.e., \verb|Para.Op1| and \verb|Para.Op2|) for communication and computation overlapping. The meaning of operations are presented in Sec~\ref{sec:planner}.


\section{Pro-Prophet Planner}\label{sec:planner}

\subsection{Lightweight Expert Placement}\label{subsec:lightweight_ep}
The design of the expert placement is crucial for efficient load balancing. For less communication of model states transferring, the planner introduces a series of lightweight expert placements.

In a lightweight expert placement, each expert is mapped to one or more devices independently. Only the parameters and gradients rather than all model states are transferred among its devices.
We use \verb|Trans| and \verb|Agg| primitives to describe these two communications respectively. In the forward pass, a \verb|Trans| is first launched to transfer the parameters. After that, each device contains the parameters of some expert, thus its local inputs routed to these experts could be computed locally. After the backward computation, the gradients of an expert could be generated in several devices. As each device only maintains the optimizer states of one expert, a \verb|Agg| primitive is launched to aggregate gradients of each expert to its original device.
This design has two advantages: 1) Only part of the model states are communicated. 2) The model states are only communicated among a subset of devices. 

\begin{figure*}[htp]
\centering
\subfloat[Traditional expert placement.]
{\label{subfig:load_EP}\includegraphics[width=0.46\linewidth]{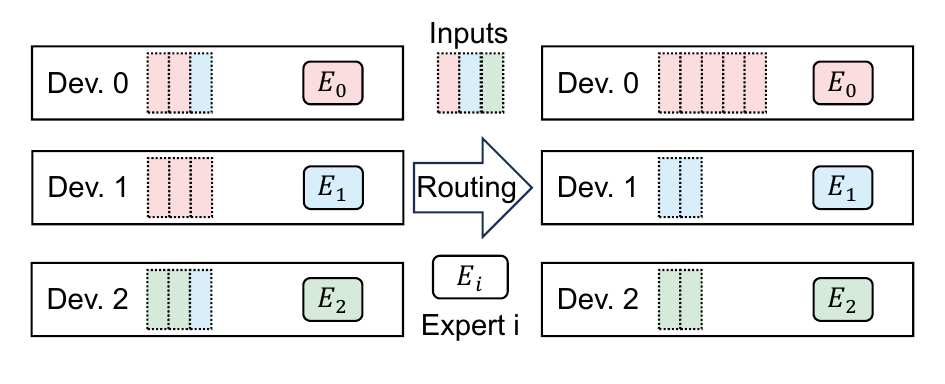}}
%
\hspace{12mm}
\subfloat[Lightweight expert placement.]{
\label{subfig:load_lightweight}
\includegraphics[width=0.46\linewidth]{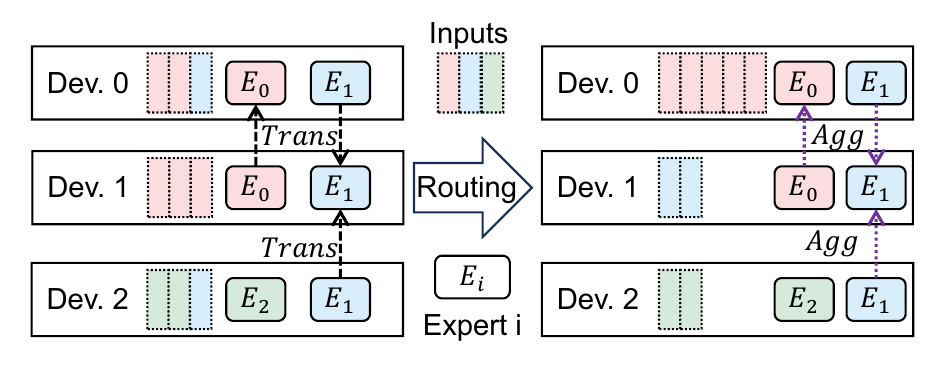}}
\caption{The comparison of a traditional and lightweight expert placement. The load is imbalanced in traditional expert placement. In a lightweight one, each expert is placed into necessary devices to balance the load. The $Trans$ and $Agg$  primitives are involved to communicate their parameters and gradients respectively.}
\label{fig:expert_placement_compare}
\end{figure*}

Fig~\ref{fig:expert_placement_compare} illustrates a comparison of a traditional and lightweight expert placement.
As shown in Fig.~\ref{subfig:load_EP}, 5, 2, and 2 inputs are routed to $E_0$, $E_1$, and $E_2$ respectively. After the A2A communication, three devices are responsible for the computation of 5, 2, and 2 inputs as each of the devices only contains parameters of a distinct expert (e.g., Dev. 0 contains $E_0$'s parameters), resulting in an imbalanced load among devices. Fig.~\ref{subfig:load_lightweight} shows a balanced load achieved by the lightweight expert placement. Experts are mapped to devices according to the routing results produced by the gate network. Parameters of $E_0$ are sent from Dev. 0 only to Dev. 1 as inputs in Dev. 2 are not routed to $E_0$. Similarly, parameters of $E_1$ are transferred to Dev. 0 and Dev. 1 for their expert computation. It maps experts to necessary devices and only communicates their parameters and gradients, effectively avoiding heavy model states transferring.

\subsection{Performance model}\label{subsec:cost_model}
It's necessary to evaluate lightweight expert placements under various device loads. 
Therefore, the planner abstracts a performance model to estimate the execution time of a MoE layer employing a lightweight expert placement. Table~\ref{tab:notation} presents notations and descriptions used in the performance model. 

\begin{table}[tb]
\caption{Notations}
\centering
    \scalebox{0.9}{
    \begin{tabular}{c|c}
    \hline
       Notation  &  Description \\
    \hline
        $T$ & Execution time of an operation \\
        $R$ & Inputs received by a device from other devices \\
        $\overline{B}$ & Average communication bandwidth \\
        $H$ & Inputs computed in a device \\
        $t$ & Computation throughput \\
        $s$ & Number of selected experts should be transferred \\
        $n$ & Number of devices a selected expert not be transferred to \\
        $E$ & Number of experts in a MoE layer\\
        $D$ & Number of devices \\  
    \hline
    \end{tabular}
    }

\label{tab:notation}
\end{table}

After employing a lightweight expert placement, a MoE layer performs four A2A communication operations, one forward expert computation operation \verb|EFC|, one backward computation operation \verb|EBC|, one \verb|Trans| operation, and one \verb|Agg| operation. To accurately evaluate the execution time of the MoE layer, we establish our performance model according to the implementation of operations and hardware characteristics.

\textbf{A2A communication.}
Tutel\cite{tutel} presents an efficient A2A implementation used in the training of a MoE model. In this implementation, devices use point-to-point(P2P) communication primitives to achieve the A2A communication operation. 
Based on this, we define the execution time of an A2A operation as below.
\begin{equation}
T_{A2A}(R) = \mathop{\max}\limits_{i}\frac{R_i \cdot size(input)}{\overline{B}}\label{eq:A2A} ,
\end{equation}
where $R_{i}$ is the total number of inputs received by device-$i$ from other devices and $size(input)$ is the size of a input.

\textbf{Expert computation.}
Next, we formulate the duration of the forward and backward expert computation.
In the expert computation procedure, the computations of devices are performed simultaneously. However, computations of different experts are launched sequentially in a device. To depict this characteristic, we define the execution time of \verb|FEC| as

\begin{equation}
T_{FEC}(H) = \mathop{\max}\limits_{i}{\frac{H_{i}}{t}}\label{eq:MoE_Fcomp}, 
\end{equation}
where $H_{i}$ is the number of inputs computed in device-$i$.

It is widely recognized that the time required for backward computation in DNN training is roughly double that of forward computation, which is the same for MoE model training. Therefore, we define the execution time of \verb|BEC| as
\begin{equation}
T_{BEC}(H) = 2\mathop{\max}\limits_{i}{\frac{H_{i}}{t}}\label{eq:MoE_Bcomp}, 
\end{equation}

\textbf{Trans and Agg primitives.}
Finally, we formulate the overhead of \verb|Trans| and \verb|Agg| primitives. 
The duration time of \verb|Trans| and \verb|Agg| primitives depends on two elements. The first element is the number of transferred experts, which determines communication rounds. The second element is the number of devices communicated in a primitive, which influences the communication scales.
Therefore, the $T_{Trans}(s, n)$ and $T_{Agg}(s, n)$ are defined as below.
\begin{equation}\label{eq:Trans}
T_{Trans}(s, n) = \frac{s * (D-n) * size(e_j.params)}{D * \overline{B}} ,
\end{equation} 
\begin{equation}\label{eq:Ag}
T_{Agg}(s, n) = \frac{s * (D-n) * size(e_j.grads)}{D * \overline{B}} ,
\end{equation}
where the $size(e_j.params)$ and $size(e_j.grad)$ are the size of parameters and gradients for the $j$-th expert.  

In summary, the overall execution time of the MoE layer with lightweight expert placement can be represented as

\begin{equation}
\label{eq7}
\begin{split}
T'(R, H, s, n) = &4T_{A2A}(R) + 3T_{FEC}(H) \\
          &+ T_{Trans}(s, n) + T_{Agg}(s, n) \\ 
\end{split}
\end{equation}

\subsection{Locality-based Greedy Algorithm}\label{subsec:search_alg}

The performance model can accurately estimate the execution time of a MoE layer deploying any expert placements. However, it is necessary to determine a communication-efficient one in various load imbalance scenarios. There are $2^{N*E}$ potential lightweight expert placements. The brute force search algorithm is time-consuming and could be a performance bottleneck. 

Therefore, the planner offers an efficient greedy search algorithm shown in Algorithm~\ref{alg1}. Taking the results of gate network $gating$, $s$ and $n$ as input, Algorithm~\ref{alg1} iteratively generates and evaluates for a better expert placement until the load is balanced. Finally, it outputs a communication-efficient expert placement $PoE$. 

Initially, the algorithm estimates the execution time of a MoE layer without implementing any lightweight expert placements and records it as minimum time. Then it employs two greedy strategies to generate a lightweight expert placement that optimizes the load of devices. Specifically, it prioritizes the expert with the higher number of responsible inputs for selection and transfers its parameters to devices that hold more inputs processed by the expert. The algorithm maintains a list of $L$ and $n\_bottoms$ to record the expert placement. Then the algorithm evaluates the expert placement using the performance model. It updates the minimum time and a counter if the current expert placement achieves a better performance. 
The search process is repeated until the load is imbalanced. The condition of the balanced load is 
\begin{equation}
\label{eq:balanced}
\max(H) - \min(H) < \alpha \frac{I}{E}, 
\end{equation}
where $I$ is the number of inputs training in an iteration and $\alpha$ is a regulable coefficient for different requirements of load balance.

As the search algorithm is required to run during the MoE model training, we define a primitive \verb|Plan| to describe this search process. 
As mentioned in Sec.~\ref{sec:back_moti}, the input distributions of adjacent iterations are similar, which inspired us to predict the distribution and reduce the frequency of execution of the algorithm. Based on the inspiration, the planner upgrades the algorithm to a locality-based one. Users can adjust the frequency of the search algorithm flexibly for better training efficiency.

\RestyleAlgo{ruled}
\SetKwComment{Comment}{/* }{ */}
\begin{algorithm}
\caption{Greedy search algorithm}\label{alg1} 

\KwIn{Inputs-to-experts mapping\ $gating$}  
\KwIn{$n$}
\KwResult{Communition-efficient expert placement $PoE$} 

\tcp{Preliminary}

$T_{output} \gets T'(R, H, 0, 0)$\;

$H, R \gets GetH\&R(gating)$\;

$L, n\_bottoms \gets [], []$\;

$cnt \gets 0$\;

\tcp{Iteratively search}

\While{$not\ balanced$}{

    \tcp{Get the index of the heaviest device} 

    $i \gets \mathop{\arg\max}\limits_{i}(H)$\; 
    
    \If{$i\  in\ Used$}{
        break;
    }
    
    $Used.append(i)$\;

    \tcp{Determine $n$ devices saving the smallest number of inputs for expert-$i$}
    $n\_bottom \gets BottomK(gating, n)$ \;
    
    $L.append(i)$\;
    
    $n\_bottoms.append(n\_bottom)$

    $s \gets size(L)$
    
    \tcp{Replace inputs among devices according to the expert placement}
    $H, R \gets Replace\_Inputs(L, n\_bottoms)$

    \tcp{Evaluate the expert placement}
    $T_{changed} \gets T'(R, H, s, n)$\;
    
    \If{$T_{changed} < T_{output} $}{    
        $T_{output} \gets T_{changed}$\;
        $cnt = s\;$
    }
    
}

\tcp{Return the communication-efficient expert placement}
$PoE \gets Get\_PoE(L[0:cnt], n\_bottoms[0:cnt])$

RETURN $PoE$;

\end{algorithm}

\section{Pro-Prophet Scheduler}\label{sec:scheduler}
Previous works introduce a search process (corresponding to \verb|Plan| primitive), model states transferring (corresponding to \verb|Trans| and \verb|Agg| primitives) to balance the load. However, their execution is blocked by other operators due to data dependency, constraining further improvement of training efficiency. In this section, we introduce designs of the scheduler which extensively overlap computation and communication based on the locality described in Sec.~\ref{sec:back_moti}. 

\subsection{Scheduling space establishment}

\begin{figure}[tbp]
\centerline{\includegraphics[width=1.0\linewidth]{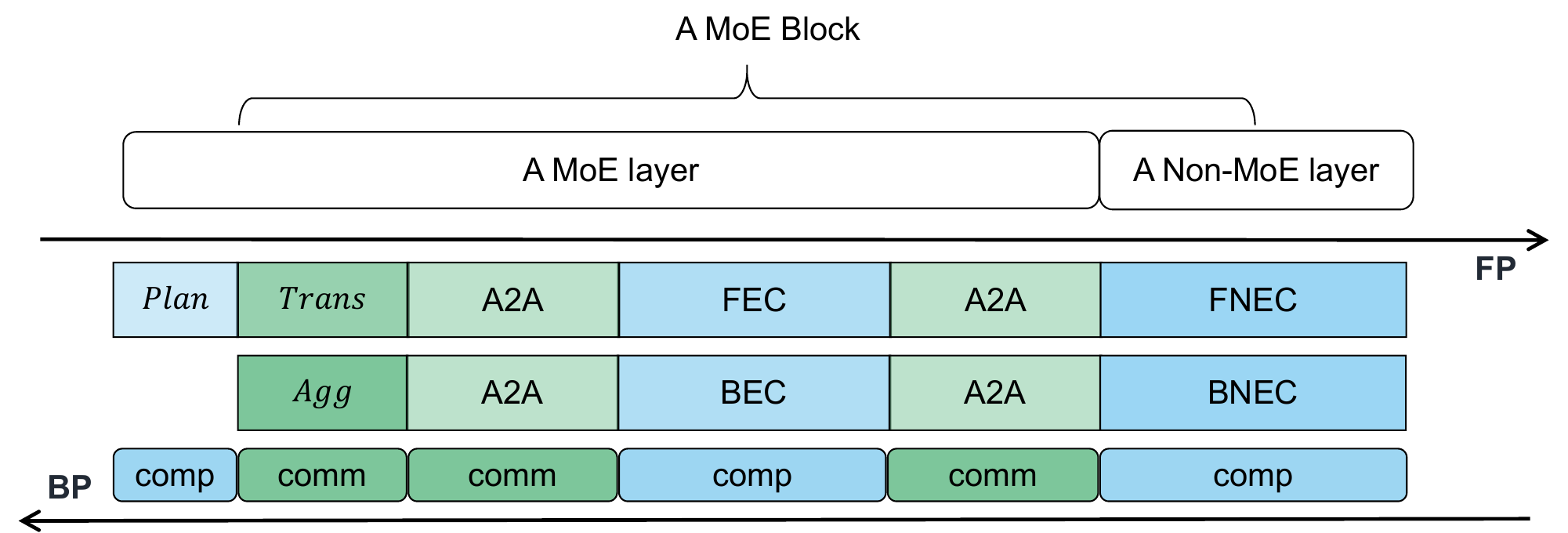}}
\caption{The device state of operators in a MoE block. FEC and BEC
are forward and backward computations of the MoE layer. FNEC and BNEC are forward and backward computations of the non-MoE layer. 
An operator is marked as \textit{comm} and a green rectangle if devices only communicate during its execution. Similarly, we use \textit{comp} and blue rectangle to mark computation operators.}
\label{fig:device_idle}
\end{figure}


As shown in Fig.~\ref{fig:MoE_structure}, a MoE model consists of both MoE and non-MoE layers stacked on top of each other. We combine every adjacent pair of MoE layer and non-MoE layer into a MoE block. 

Fig.~\ref{fig:device_idle} presents the device state of operators in a MoE block. Only primary operations are presented for clarity description. 
During the FP, two primitives related to load balancing(i.e., \verb|Plan| and \verb|Trans|) and three basic operations(i.e., two A2A communications and a forward expert computation \verb|FEC|) are performed in a MoE layer. For the non-MoE layer, only a forward computation \verb|FNEC| is executed. After the gate network produces the routing decision and input distribution, a \verb|Plan| operator is performed to identify a load-balancing strategy based on the load of devices. The \verb|Trans| primitive can only be launched once the strategy is determined. And then, two A2A communications and an expert computation can be launched.
During the BP, the MoE layer executes a \verb|Agg| operation, two A2A communications, and a backward expert computation \verb|BEC|. Only a backward computation \verb|BNEC| is performed in the non-MoE layer, the \verb|Agg| operation is carried out to aggregate the gradient of experts following the completion of their backward computation. 

We label an operation as \textit{comm} if devices communicate during the execution of this operation. Similarly, if an operation computes all the time, we label it as \textit{comp}. 
Before the launching of \verb|Plan|, information related to produce a load-balancing strategy is stored in each device. Therefore, there are only computations during \verb|Plan|. We label it as \textit{comp}. As \verb|Trans| primitive only exchanges the parameters of experts based on the load balancing strategy, we label it as \textit{comm}. According to the description of Sec.~\ref{sec:back_moti}, the A2A is marked as \textit{comm}, and the computation of the MoE layer and non-MoE layer are tagged as \textit{comp}.
As for \verb|Agg| primitive, gradients of the same expert are aggregated into a device. We flag it as \textit{comm}.

The operators with data dependency are tightly interconnected along the timeline, constraining the scheduling of computation and communication. 
However, the locality mentioned in Sec.~\ref{sec:back_moti} enables the pre-launch of some data-dependent operations without breaking the data dependency. 
We can insert the data-independent operation into data-dependent operators to make room for communication and computation overlapping. Specifically, in the case of a MoE layer, the input distribution for the current iteration can be estimated by leveraging the distribution from former iterations, which allows us to produce the load-balancing strategy by invoking a \verb|plan| operator within earlier iterations.
Subsequently, \verb|Trans| primitives can be scheduled to earlier locations on the timeline. We also find that the \verb|Agg| primitive is independent of computation operations later, thus we can schedule it to later positions.

\begin{figure*}[ht]
\centerline{\includegraphics[width=1\linewidth]{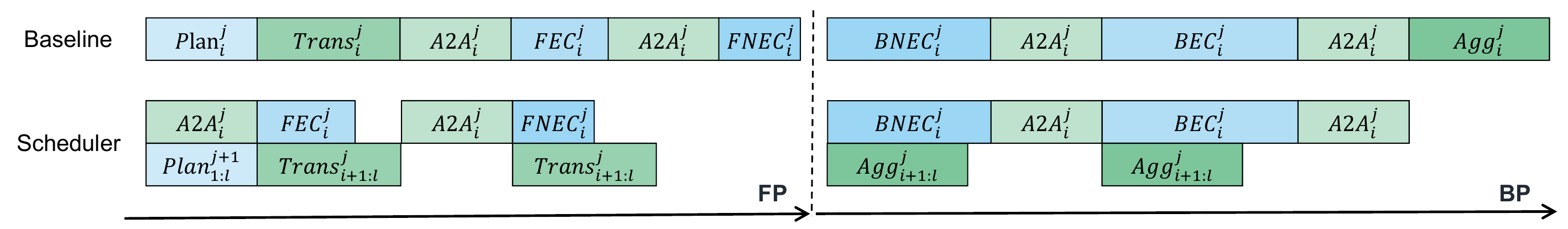}}
\caption{Scheduling space. The subscript denotes the index of the MoE block and the superscript denotes that of the iteration. For example, $Trans_{j}^{i+1:l}$ denotes the set of $Trans$ operations spanning from the block $i+1$ to the block $l$ during the iteration $j$, where $l$ is the number of MoE blocks.}

\label{fig:overall_scheduling}
\end{figure*}
The analysis above provides the potential for computation and communication scheduling. However, there are several constraints for arbitrarily scheduling. Firstly, the estimation of the distribution means that we can establish a load-balancing strategy within a former iteration. As the distribution of the last iteration is necessary for higher estimation accuracy, the earliest position of a \verb|plan| primitive in $i$-th iteration is the $i-1$-th iteration. Secondly, there are two main ways to update parameters. It is necessary to update the expert parameters before the \verb|Trans| primitive. We can perform the updating procedure layer by layer\cite{li2023accelerating} or update at the end of the BP\cite{fastermoe}. For the layer-by-layer updating, we can launch the \verb|Trans| primitive of a MoE layer of an iteration within the last iteration, which does not apply to concentrated updating works. For the concentrated updating, the \verb|Trans| primitive could be performed at the end of the BP of the last iteration, which has a similar effect as starting it within this iteration.
For the universality of our method, we confine the scheduling of the \verb|Trans| primitive within a single iteration. Lastly, it is necessary to aggregate the gradients of experts at each iteration. Therefore, the replacing of the \verb|Agg| primitive is also confined within a single iteration.   

Fig.~\ref{fig:overall_scheduling} illustrates our scheduling space. The subscript denotes the index of the MoE block of the operation and the superscript denotes its iteration.
All \verb|Plan| computations of iteration $j+1$ can be scheduled to the A2A communication of iteration $j$. The \verb|Trans| primitives from block $i+1$ to $l$ during the $j$-th iteration are overlapped with the forward computations of the $i$-th block, where $l$ is the total number of blocks. 
Similarly, the \verb|Agg| primitives from block $i+1$ to $l$ are orchestrated to overlap with backward computations of the $i$-th block. 
Our scheduling space considers a MoE block as a unified reordering entity, thus overcoming the limitations in observation imposed by previous methods.

\subsection{Block-wise scheduling strategy}

\begin{figure}[htp]
\centering
\subfloat[Scheduling $Trans$ to the forward expert computation.]{
\label{subfig:trans1}
\includegraphics[width=1.0\linewidth]{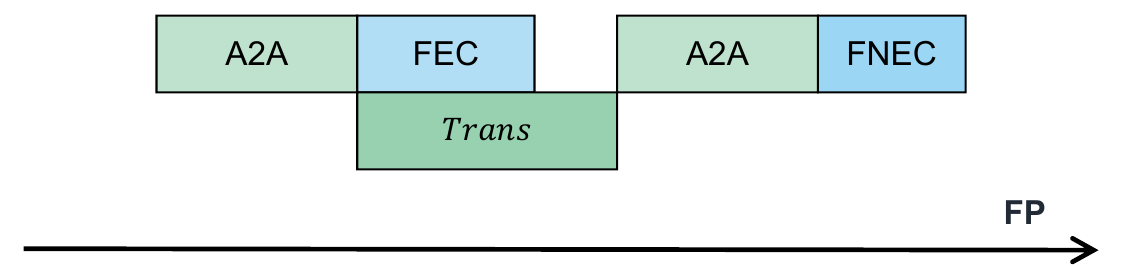}}
\quad  
\subfloat[Scheduling $Trans$ to the forward non-expert computation.]{
\label{subfig:trans2}
\includegraphics[width=1.0\linewidth]{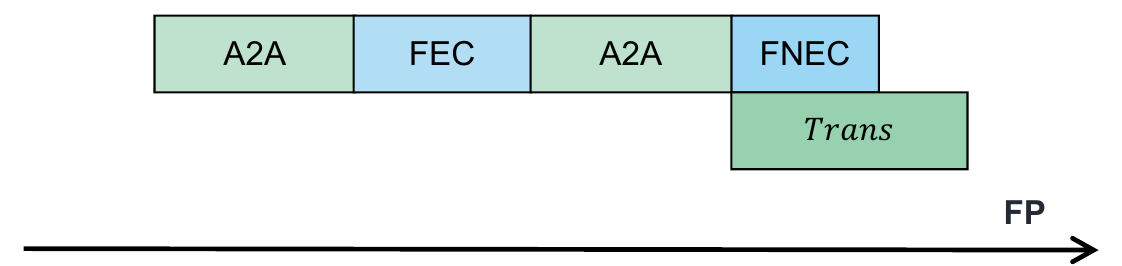}}
\quad  
\subfloat[Splitting and scheduling $Trans$ to both the forward expert computation and non-expert computation.]{
\label{subfig:trans3}
\includegraphics[width=1.0\linewidth]{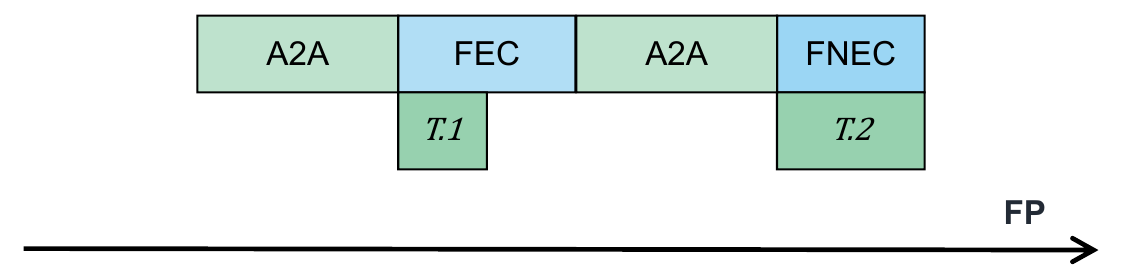}}
\caption{Different scheduling strategies for a $Trans$ primitive.}
\label{fig:Trans_scheduling}
\end{figure}

The scheduling space provides extensive strategies for scheduling. However, operator-grained scheduling is far from making full use of overlapping space. We partition the operator into sub-operators and take a scheduling at the sub-operator level. We take a brief example to describe the advantage of sub-operator scheduling in Fig.~\ref{fig:Trans_scheduling} which shows three types of scheduling for a \verb|Trans| primitive. The overhead of \verb|Trans| primitive is varied as the load of devices (e.g., the number of heavy-load experts changed as the training). As shown in Fig.~\ref{subfig:trans1} and Fig.~\ref{subfig:trans2},  
a forward computation of a MoE layer or a non-MoE layer cannot afford the hiding of a \verb|Trans| primitive as their short duration time. Consequently, the \verb|Trans| primitive will block the process of the model training. Fig.~\ref{subfig:trans3} presents the sub-operators scheduling. The \verb|Trans| primitive is split into two sub-operators scheduled to two computations respectively. The sub-operator scheduling improves the utilization of overlapping space and reduces the communication overhead. 

A dynamic scheduling strategy is beneficial as the load of devices fluctuates with the training process. However, the non-negligible overhead will be introduced if we determine an optimal scheduling strategy that as far as possible hides the overhead of three types of primitives involved by systematical load balancing methods in the runtime. 
Therefore, we design an offline scheduling policy to overlap communication and computation while avoiding the extra overhead. The policy is founded on static elements within dynamics, not intuition. 
Our block-wise scheduling strategy is summarized in Algorithm~\ref{alg:offline}. 
The first primitive is \verb|Plan|. We schedule the \verb|Plan| primitive of the $i$-th block in the iteration $j+1$ to the A2A communication of the $i$-th block in the $j$-th iteration. 
For the \verb|Trans| primitive, we overlap it with computations of the former block within an iteration. Specifically, the forward computation of the $i$-th block is responsible for overlapping the \verb|Trans| primitive of block $i+1$. As two computations are executed in the $i$-th block, we split the \verb|Trans| primitive into two sub-primitives and launch them simultaneously with the two computations. Even though the duration of \verb|Trans| and \verb|EFC| varies as the device loads, the forward computation overhead of the non-MoE layer and the transferring overhead of an expert's parameters are static. We can estimate them before training and properly split the \verb|Trans| primitive. The advantage of the estimation is that we can exhaustively fill in the communication idle in the performing of the forward computation of the non-MoE layer.
Finally, we schedule the $Agg$ of block $i+1$ into the backward computation of the $i$-th one. Similarly, we can estimate the backward computation overhead of the non-MoE layer and do a suitable communication partition. 


\RestyleAlgo{ruled}
\SetKwComment{Comment}{/* }{ */}
\begin{algorithm}[tb]
\caption{Block-wise scheduling strategy}\label{alg:offline} 
\KwResult{Training workflow after the scheduling.}

\For{$j$ in iterations}{
    \For{$i$ in MoE blocks}{
        \tcp{Forward propagation}

        $SubTrans1, Subtrans2$ = Partition($Trans$);

        $SubAgg1, SubAgg2$ = Partition($Agg$);

        Launch for parallel \{$A2A_i^j$, $Plan_i^{j+1}$\};
        
        Launch for parallel\{$SubTrans1_{i+1}^{j}$, $FEC_i^j$\};
        
        Launch $A2A_i^j$;

        Launch for parallel\{$SubTrans2_{i+1}^{j}$, $FNEC_i^j$\};

        \tcp{Backward propagation}

        Launch for parallel\{$SubAgg1_{i+1}^{j}$, $BNEC_i^j$\};
        
        Launch $A2A_i^j$;
        
        Launch for parallel\{$SubAgg2_{i+1}^{j}$, $BEC_i^j$\};

        Launch $A2A_i^j$;
    } 
}
\end{algorithm}

\subsection{Effective collaboration with planner}
To better integrate the planner and scheduler, we combine the scheduling performed by the scheduler into the performance model of the planner.

Specifically, we define the parallel execution time of the $Trans$ and $Agg$ primitives as $T_{PTrans}(H, s, n)$ and $T_{PAgg}(H, s, n)$ respectively. 
Besides, we denote $T_{FNEC}$ and $T_{BNEC}$ as the execution time of \verb|FNEC| and \verb|BNEC|.  
If $T_{Trans}(s, n)$ can be hidden by $T_{FEC}(H)$ and $T_{FNEC}$, then $T_{PTrans}(H, s, n)$ is equal to 0. Otherwise, $T_{PTrans}(H, s, n)$ is equal to $T_{Trans}(s, n) - T_{FEC}(H) - T_{FNEC}$. That means that $T_{PTrans}(H, s, n) = \max(0, T_{Trans}(s, n) - T_{FEC}(H) - T_{FNEC})$.
Similarly, $T_{PAgg}(H, s, n)$ can be expressed as $\max(0, T_{Agg}(s, n) - T_{BEC}(H) - T_{BNEC})$.

With the above analysis, the overall execution time of the MoE layer estimated by the performance model of the planner is changed as below:
\begin{equation}
\label{eq8}
\begin{split}
T'(R, H, s, n) = &4T_{A2A}(R) + 3T_{FEC}(H) \\
          &+ T_{PTrans}(H, s, n)+ T_{PAgg}(H, s, n).
\end{split}
\end{equation}

By combining the planner and scheduler, we can achieve a fine-grained pre-allocation of hardware resources to experts, efficiently addressing the load imbalance problem during training.

\section{Evaluation}

\textbf{Testbed.} We test Pro-Prophet on three types of nodes named \textit{HPWNV}, \textit{HPNV} and \textit{LPWNV} respectively. Each \textit{HPWNV} node is equipped with 2 Intel Xeon CPUs (2.40GHz) and 4 NVIDIA 3090 GPUs with 24GB graphics memory. Each CPU is connected to two GPUs through PCI-Express 3.0. 100Gb/s Infiniband is used for inter-node communication.
The difference between \textit{HPWNV} and \textit{HPNV} is that GPUs within a \textit{HPNV} node are connected by NVLink-3.0 connections. Specifically, four GPUs are divided into two groups. Two GPUs within a group are connected by a NVLink connection. The difference between \textit{HPWNV} and \textit{LPWNV} is that the type of GPUs of a \textit{LPWNV} node is 2080Ti. 

\textbf{Models and baselines.} As shown in table~\ref{tab:model}, we use five variants of MoE-GPT models in our experiments. All FFN layers are replaced by a MoE layer. The number of experts within a MoE layer is consistent with the number of GPUs.

We compared Pro-Prophet with two representative MoE training systems: 1) Deepspeed-MoE~\cite{deepspeed}: Deepspeed-MoE is an efficient MoE framework developed by Microsoft. It exclusively implements EP.
2) FasterMoE~\cite{fastermoe}: This training system employs a systematic load balancing method, \textit{dynamic shadowing}, to effectively accelerate the model training. 

\begin{table}[b]
\caption{Model configuration.}
\centering
\scalebox{1.2}{
    \begin{tabular}{|c|c|c|c|c|}
    \hline
        \multicolumn{2}{|c|}{Name} & Layers & Embedding & Hidden \\
    \hline
        \multicolumn{2}{|c|}{MoE-GPT-S} & 12 & 512 & 1024 \\
        \multicolumn{2}{|c|}{MoE-GPT-M} & 12 & 1024 & 2048 \\
        \multicolumn{2}{|c|}{MoE-GPT-L} & 12 & 2048 & 4096 \\
    \hline
        \multicolumn{2}{|c|}{MoE-GPT-DS} & 24 & 512 & 1024 \\
        \multicolumn{2}{|c|}{MoE-GPT-DM} & 24 & 1024 & 2048 \\
    \hline
    \end{tabular}
}
\label{tab:model}
\end{table}

\textbf{Default settings.} Unless otherwise specified, we fix some training settings.
We train MoE models on the cluster consisting of \textit{HPWNV} nodes. We evaluate Pro-Prophet within the first 100 iterations as the input distribution tends to stabilize with the training process. 

\subsection{End-to-end Performance}

In summary, Pro-Prophet achieves 1.36-2.66x and 1.01-1.48x speedups compared to Deepspeed-MoE and FasterMoE respectively.

\textbf{Experiments on \textit{HPWNV}.} We first evaluate the end-to-end performance of Pro-Prophet on two \textit{HPWNV} clusters. They contain 4 and 8 \textit{HPWNV} nodes respectively. We fixed the number of tokens trained in an iteration to 16384 and 32768.

As shown in previous works, the $k$ is set to 1 or 2\cite{gshard, switch, janus23} for better balancing between the model quality and training efficiency.
We conducted experiments under both values to validate the generality of Pro-Prophet.

\begin{figure*}[htp]
\centering
\subfloat[\label{fig:3090-16-1}4\textit{HPWNV} nodes, $k$=1, 1.01--1.23x]{
\includegraphics[scale=0.26]{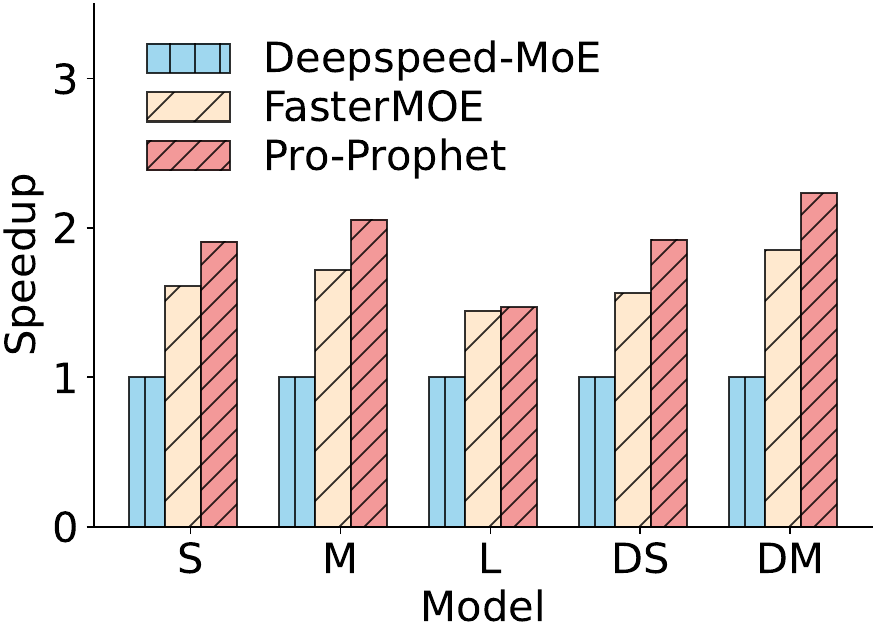}}
\quad  
\subfloat[\label{fig:3090-32-1}8\textit{HPWNV} nodes, $k$=1, 1.08--1.31x]{
\includegraphics[scale=0.26]{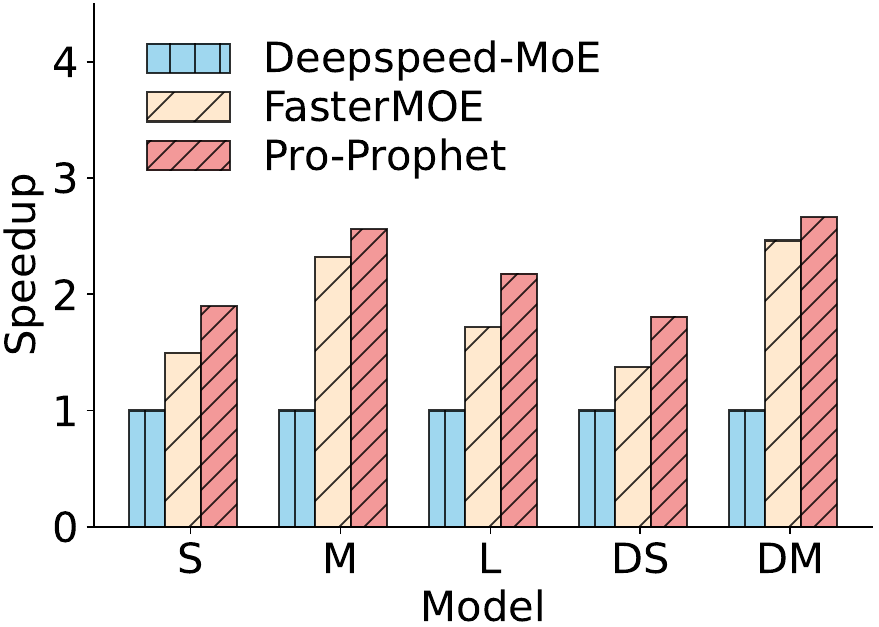}}
\quad  
\subfloat[\label{fig:3090-16-2}4\textit{HPWNV} nodes, $k$=2, 1.14--1.48x]{
\includegraphics[scale=0.26]{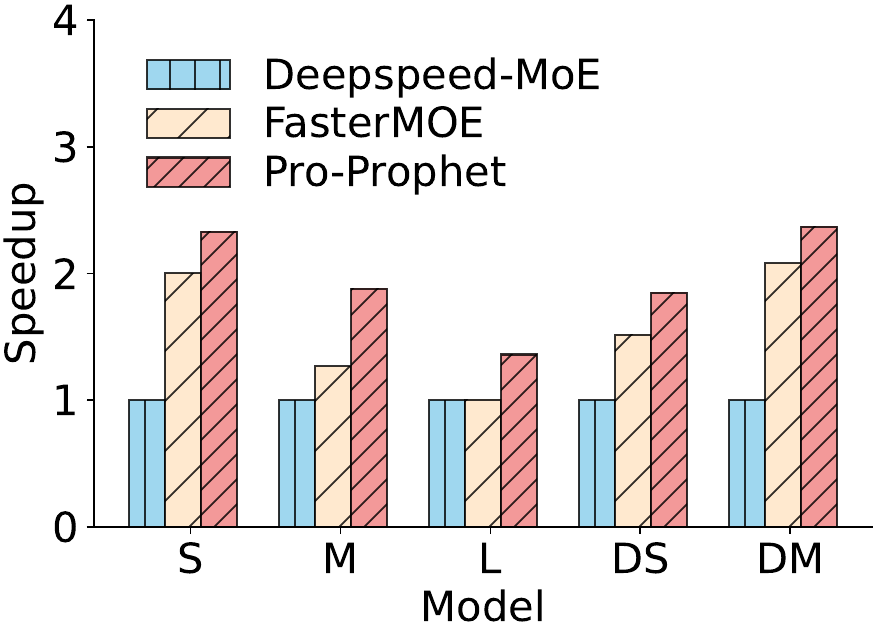}}
\quad  
\subfloat[\label{fig:3090-32-2}8\textit{HPWNV} nodes, $k$=2, 1.05--1.31x]{
\includegraphics[scale=0.26]{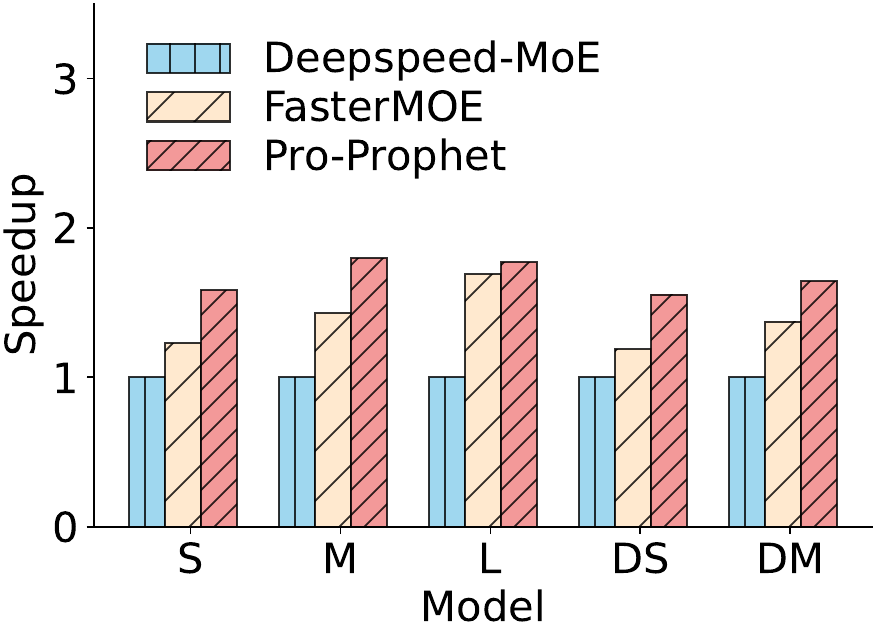}}

\caption{End-to-end performance. The numbers in the captions denote speedups achieved by Pro-Prophet over the best baseline.}
\label{fig:exp-3090}
\end{figure*}

Fig.~\ref{fig:3090-16-1} and Fig.~\ref{fig:3090-32-1} illustrate speedups achieved by Pro-Prophet under five benchmark models with a top-$1$ gate network.
Pro-Prophet achieved 1.47-2.66x end-to-end performance gains in comparison with Deepspeed-MoE. As to FasterMoE, Pro-Prophet achieves performance enhancements of up to 1.31x with an average of 1.19x. 



As shown in Fig.~\ref{fig:3090-16-2} and Fig.~\ref{fig:3090-32-2}, speedups under a top-$2$ gate network achieved by Pro-Prophet are 1.36-2.37x and 1.05-1.48x compared to Deepspeed-MoE and FasterMoE respectively.
The result shows the coarse-grained and blocked manner of FasterMoE will introduce additional runtime overhead and hinder the further improvement of the training efficiency. Our method can precisely pre-allocate resources to experts, thereby avoiding this issue.


\begin{table}[tb]
\renewcommand\arraystretch{1.5}
\caption{The overall speedup on 4 \textit{HPNV} nodes.}
\centering
\scalebox{0.9}{
    \begin{threeparttable}
    \begin{tabular}{|c|c|c|c|c|c|}
    \hline
        \multirow{2}*{K} & \multirow{2}*{GPUs} & \multirow{2}*{Tokens} & \multirow{2}*{Model} & \multicolumn{2}{c|}{Speedup to DeepspeedMoE} \\
        \cline{5-6}
        ~ & ~ & ~ & ~ & FasterMoE & Pro-Prophet \\
    \hline
        \multirow{5}*{1} & \multirow{10}*{16} & \multirow{10}*{16384} & MoE-GPT-S & 1.63 & \textbf{1.98} \\ \cline{4-6} 
        ~ & ~ & ~ & MoE-GPT-M & 1.99 & \textbf{2.22} \\ \cline{4-6} 
        ~ & ~ & ~ & MoE-GPT-L & 1.62 & \textbf{1.80} \\ \cline{4-6} 
        ~ & ~ & ~ & MoE-GPT-DS & 1.34 & \textbf{1.70} \\ \cline{4-6} 
        ~ & ~ & ~ & MoE-GPT-DM & 1.68 & \textbf{2.26} \\ \cline{1-1} \cline{4-6}
        \multirow{5}*{2} & ~ & ~ & MoE-GPT-S & 2.31 & \textbf{2.62} \\ \cline{4-6} 
        ~ & ~ & ~ & MoE-GPT-M & 1.82 & \textbf{2.10} \\ \cline{4-6} 
        ~ & ~ & ~ & MoE-GPT-L & 1.94 & \textbf{2.23} \\ \cline{4-6} 
        ~ & ~ & ~ & MoE-GPT-DS & 1.77 & \textbf{1.94} \\ \cline{4-6} 
        ~ & ~ & ~ & MoE-GPT-DM & 1.84 & \textbf{2.07} \\ \cline{4-6} 
    \hline
    \end{tabular}
        \begin{tablenotes}    
        \footnotesize              
        \item[1] The label "Token" is the number of tokens trained in an iteration.           
        \end{tablenotes}              
    \end{threeparttable}
}
\label{tab:HPNV}
\end{table}

\textbf{Experiments on HPNV and LPWNV clusters.} Different hardware conditions significantly affect the effectiveness of the method. For example, the device memory may serve as a constraint on the maximum training tokens in an iteration. Besides, the training issue may be altered by variations in computing throughput and communication bandwidth, causing a significant influence on the effectiveness of methods.
To verify the generality of Pro-Prophet, we conduct experiments on diverse hardware environments with varying memory consumption, and the rate of computing throughput and communication bandwidth. 

Due to the limited memory capacity compared to the \textit{HPWNV} and \textit{HPNV}, we only train the four smaller models listed in Table \ref{tab:model} on the \textit{HPWNV} cluster. The number of tokens trained in one iteration is set to 4096. 

\begin{table}[tb]
\renewcommand\arraystretch{1.5}
\caption{The overall speedup on a 2 \textit{LPWNV} nodes.}
\centering
\scalebox{0.9}{
    \begin{threeparttable}
    \begin{tabular}{|c|c|c|c|c|c|}
    \hline
        \multirow{2}*{K} & \multirow{2}*{GPUs} & \multirow{2}*{Tokens} & \multirow{2}*{Model} & \multicolumn{2}{c|}{Speedup to DeepspeedMoE} \\
        \cline{5-6}
        ~ & ~ & ~ & ~ & FasterMoE & Pro-Prophet \\
    \hline
        \multirow{4}*{1} & \multirow{8}*{8} & \multirow{8}*{4096} & MoE-GPT-S & 1.20 & \textbf{1.30} \\ \cline{4-6} 
        ~ & ~ & ~ & MoE-GPT-M & 1.02 & \textbf{1.18} \\ \cline{4-6} 
        ~ & ~ & ~ & MoE-GPT-DS & 1.12 & \textbf{1.30} \\ \cline{4-6} 
        ~ & ~ & ~ & MoE-GPT-DM & 0.96 & \textbf{1.26} \\ \cline{4-6} 
        \cline{1-1}
        \multirow{4}*{2} & ~ & ~ & MoE-GPT-S & 1.56 & \textbf{1.91} \\ \cline{4-6} 
        ~ & ~ & ~ & MoE-GPT-M & 1.29 & \textbf{1.94} \\ \cline{4-6} 
        ~ & ~ & ~ & MoE-GPT-DS & 1.44 & \textbf{1.64} \\ \cline{4-6} 
        ~ & ~ & ~ & MoE-GPT-DM & 1.25 & \textbf{1.58} \\ \cline{4-6} 
    \hline
    \end{tabular}
        \begin{tablenotes}    
        \footnotesize              
        \item[1] The label "Token" is the number of tokens trained in an iteration.        
        \end{tablenotes}              
    \end{threeparttable}
}
\label{tab:LPWNV}
\end{table}

\begin{figure}[htbp]
\centering
\subfloat[\label={$k$=1}]{
\includegraphics[scale=0.25]{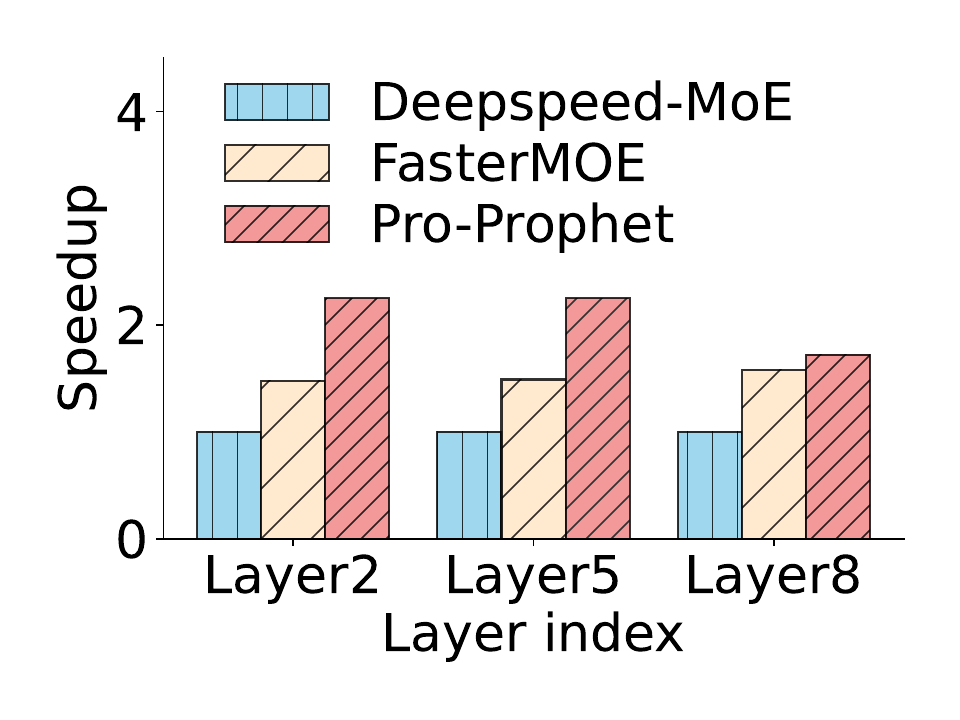}}
\quad  
\subfloat[\label={$k$=2}]{
\includegraphics[scale=0.25]{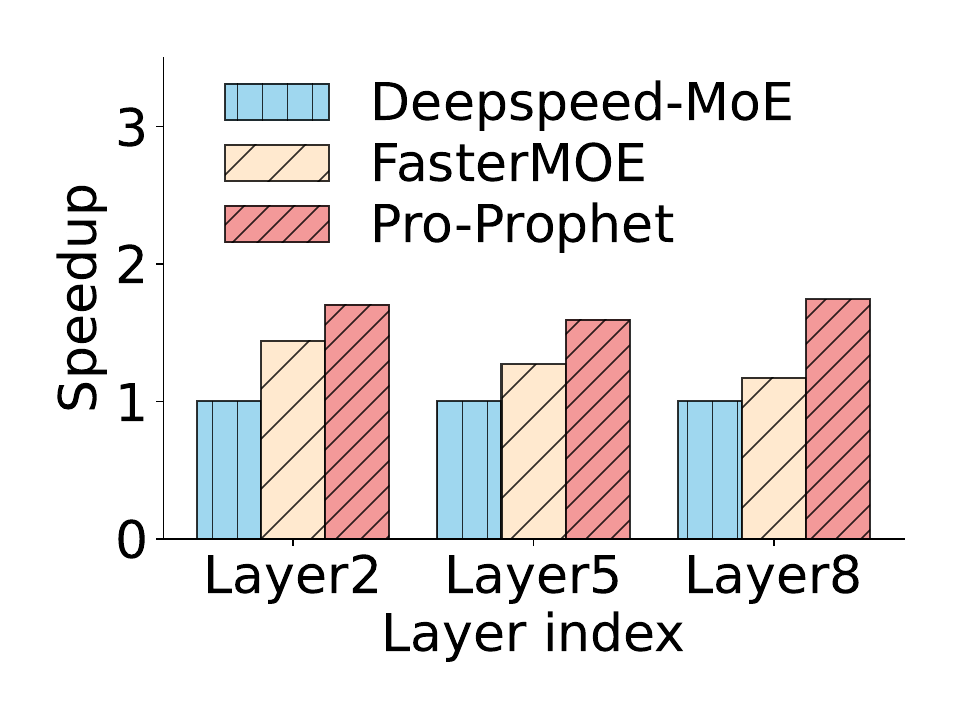}}
\caption{Speedups across different layers over the Deepspeed-MoE in the MoE-GPT-M model with different values of $k$. Pro-Prophet achieves 1.09-1.49x single-layer speedups compared to FasterMoE.}
\label{fig:layer_speedup}
\end{figure}

Table~\ref{tab:HPNV} shows the speedup results under various models and values of $k$ in a cluster consisting of 4 \textit{HPNV} nodes. The highest speedups are highlighted in the table. With the equipment of NVLink connections in the cluster, communication processes such as A2A will be accelerated. Under this condition, Pro-Prophet achieves 1.71-2.63x speedups compared to Deepspeed-MoE and 1.10-1.35x to FasterMoE, demonstrating its adaptability to conditions with higher communication bandwidth. 
We also test Pro-Prophet on a cluster consisting of 2 \textit{LPWNV} nodes. The results are presented in Table~\ref{tab:LPWNV} and the highest speedups are emphasized too.
Due to the lower computation ability of the 2080Ti compared to the 3090 GPU, the impact of the computation process becomes more significant. In this environment, Pro-Prophet achieved a speedup of 1.18-1.94x compared to Deepspeed-MoE and 1.08-1.50x compared to FasterMoE, showing its robustness in conditions with lower computation power.

The result of the MoE-GPT-DM model with $k$=1 shows that Deepspeed-MoE achieves higher performance compared to FasterMoE as FasterMoE transports parameters to unnecessary devices, resulting in additional runtime overhead. 
However, Pro-Prophet can accurately find a communication-efficient expert placement, thereby avoiding this trouble.

\subsection{Fine-grained analysis of Pro-Prophet}\label{sec:fine-grained_exp}

We conducted a fine-grained analysis of Pro-Prophet according to speedups in a single layer and a single iteration.
Experimental results demonstrate that Pro-Prophet enhances training performance in each layer and iteration during training.

\textbf{Single-layer speedup.} We first evaluate the single-layer performance of Pro-Prophet. Fig. \ref{fig:layer_speedup} illustrates the execution time across different layers of three methods on the MoE-GPT-M model. We randomly select the index of layers and use the Pytorch Profiler to collect the training time. As shown in the figure, Pro-Prophet achieves 1.60-2.25x single-layer speedups compared to Deepspeed-MoE and 1.09-1.49x to FasterMoE. 

Varying loads of experts across layers occur in the training process. This phenomenon leads to fluctuating speedups for Pro-Prophet. However, it consistently outperforms two baselines in different layers, demonstrating its superior capability of load balancing under diverse load-imbalance conditions. 


\begin{figure}[tbp]
\centerline{\includegraphics[width=0.95\linewidth]{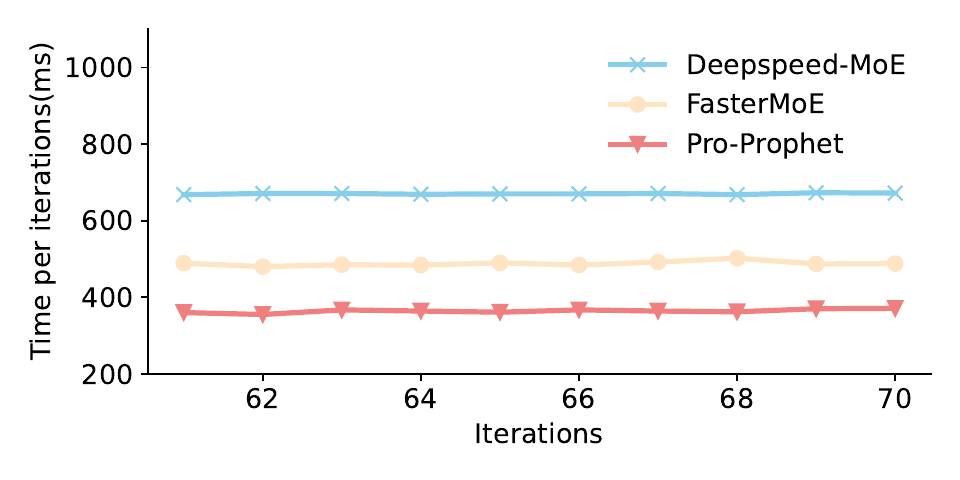}}
\caption{Per-iteration execution time in the MoE-GPT-M model when $k$=1. Pro-Prophet achieved 1.34x speedup on average compared to fasterMoE.}
\label{fig:single_iteration}
\end{figure}


\textbf{Single-iteration speedups.} 
We also evaluate the single-layer performance of Pro-Prophet. We conduct experiments on the MoE-GPT-M model with $k$=1. The results are presented in Fig. \ref{fig:single_iteration}. Compared to fasterMoE, Pro-Prophet achieved 1.34x speedup on average. The iteration time of Pro-Prophet is consistent and lower. This phenomenon can mainly be attributed to the fact that Pro-Prophet is capable of adapting to dynamic situations.

\subsection{Ablation study}\label{sec:Ablation_exp}

\textbf{Necessity of the dynamic adaptation.} Dynamic adaption is necessary for MoE models. It's reasonable to transfer heavy-load experts to other GPUs, but it's unclear if Pro-Prophet's dynamic search algorithm is necessary. 

To certify the necessity of our algorithm, we compare the planner with two simple dynamic policies. Specifically, two policies transfer 2 and 3 experts with the heaviest load to all GPUs. We named them \textit{top2} and \textit{top3} respectively. We use PyTorch's \verb|topk| function to implement these strategies. The overhead of determining the heaviest experts is negligible.

\begin{figure}[tbp]
\centering
\subfloat[\label={$k$=1}]{
\label{policy_MoE-M-k=1}
\includegraphics[scale=0.25]{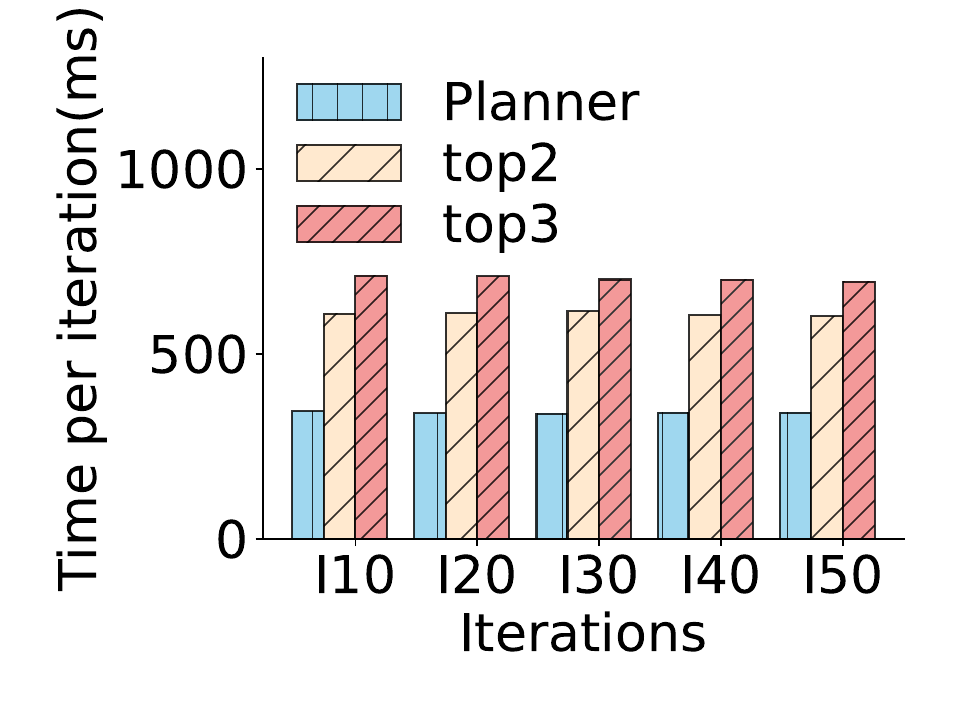}}
\quad  
\subfloat[\label={$k$=2}]{
\label{policy_MoE-M-k=2}
\includegraphics[scale=0.25]{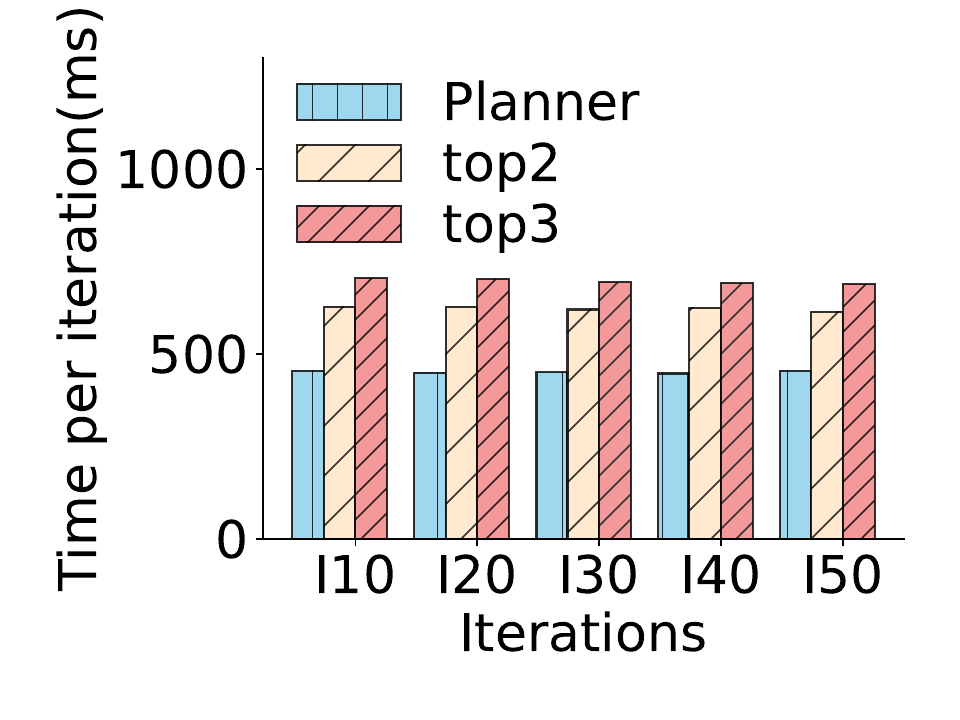}}
\caption{The iteration latency of different policies in the MoE-GPT-M model. }
\label{fig:policy_compare}
\end{figure}

Figure \ref{fig:policy_compare} illustrates the latency of three policies in a single iteration with different values of $k$. As shown in Fig. \ref{policy_MoE-M-k=1}, the planner gains 1.77-1.82x speedups compared to the \textit{top2} policy and 2.04-2.10x speedups to the \textit{top3} policy when $k$=1.
The results shown in Fig.~\ref{policy_MoE-M-k=2} demonstrate that the planner gains speedups ranging from 1.38-1.40x compared to different policies when $k$=2. 

The experimental results indicate that fixing the number of experts and passing them to all GPUs does not yield good results. The input distribution changes as training progresses, resulting in different optimal expert placements. Compared to these two dynamic strategies, our algorithm introduces more overhead, but it's necessary for faster training speeds. 

\begin{figure}[tbp]
\centerline{\includegraphics[scale=0.45]{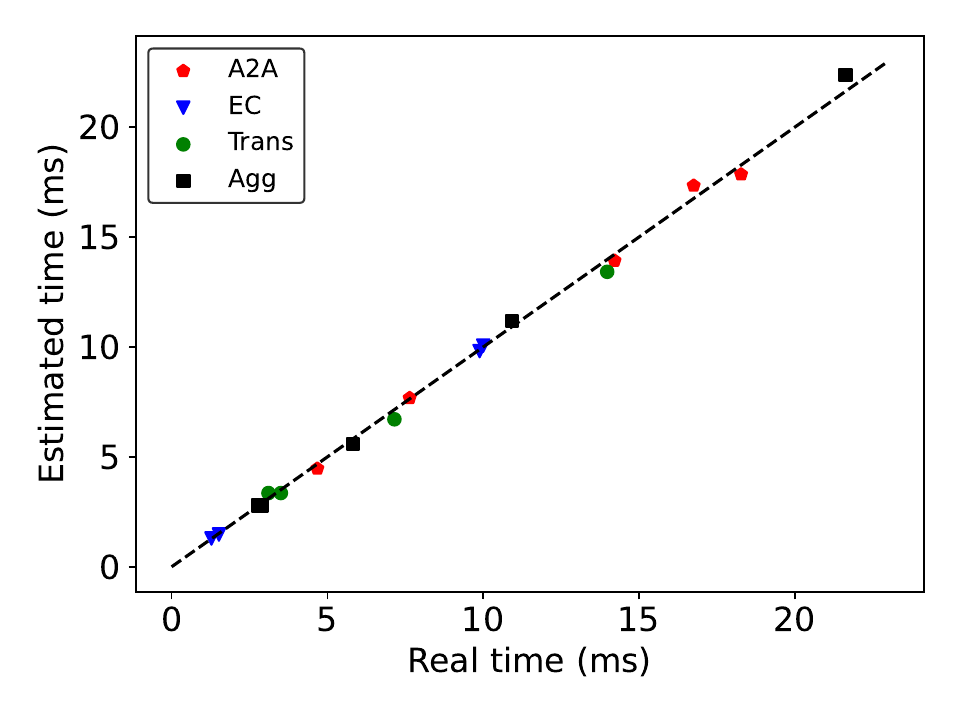}}
\caption{The accuracy of the performance model. The mean estimation error is less than 5\%.}
\label{fig:performance_model}
\end{figure}

\textbf{Accuracy of performance model.} Fig.~\ref{fig:performance_model}
illustrates the accuracy of the performance model. We compare the estimated time to the real time on A2A, expert computation (EC), \verb|Trans| and \verb|Agg| operations. The results show that our mean estimation error is less than 5\%.

\textbf{Effectiveness of components.} 
To verify the effectiveness of the components, we conduct incremental experiments. We first turn off all optimizations for Pro-Prophet and use it as a baseline. Based on this, we activate the planner and scheduler sequentially and record the speedup they attain relative to the baseline. Finally, we verify the effective combination of the planner and scheduler mentioned in Sec.~\ref{sec:scheduler}.

\begin{figure}[tbp]
\centerline{\includegraphics[scale=0.5]{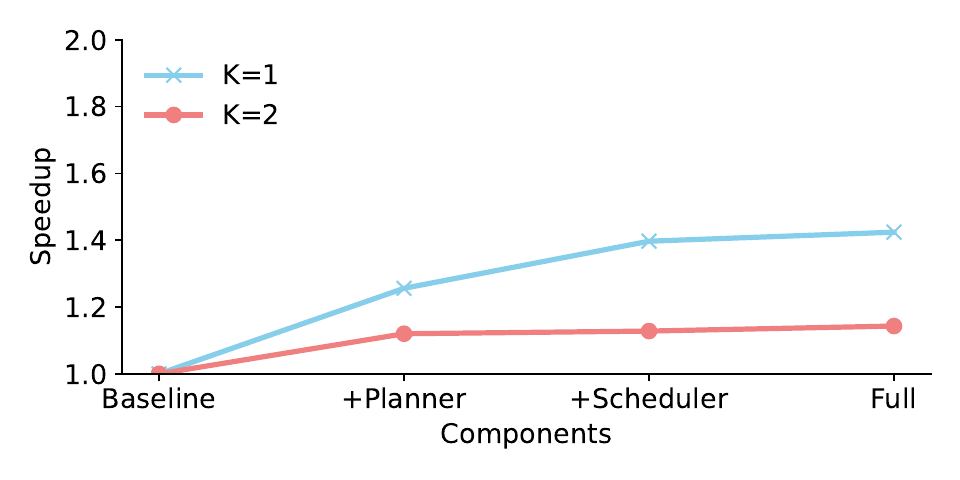}}
\caption{The effectiveness of components. The baseline is Pro-Prophet without any optimizations. Full is the condition of turning on the effective combination of the planner and scheduler. The planner, scheduler and Full achieve a speedup of 1.19x, 1.075x and 1.025x on average respectively.}
\label{fig:components}
\end{figure}

Fig.~\ref{fig:components} demonstrates speedups on the MoE-GPT-M under different $k$. Compared to the baseline, the planner
gains 1.26x and 1.12x speedups when $k$=1 and $k$=2 respectively. These results show that the planner can efficiently and accurately determine a communication-efficient expert placement for well load-balancing under different conditions.
Besides, the scheduler gains 1.14X and 1.01x speedups when $k$=1 and $k$=2 respectively. These verify that the scheduler can hide the overhead of load-balancing, further improving the training performance. The speedups achieved by the scheduler are significantly influenced by the expert placement produced by the planner. Finally, we test the effectiveness of the effective combination (\textit{Full} in the figure). As the performance model estimates the overlapped execution time, the planner will further balance the load based on the scheduler's capability of communication and computation overlapping.
The results demonstrate that it achieves 1.03x and 1.02x speedups under different values of $k$.

\begin{figure}[tbp]
\centering
\subfloat[\label={$k$=1}]{
\includegraphics[scale=0.25]{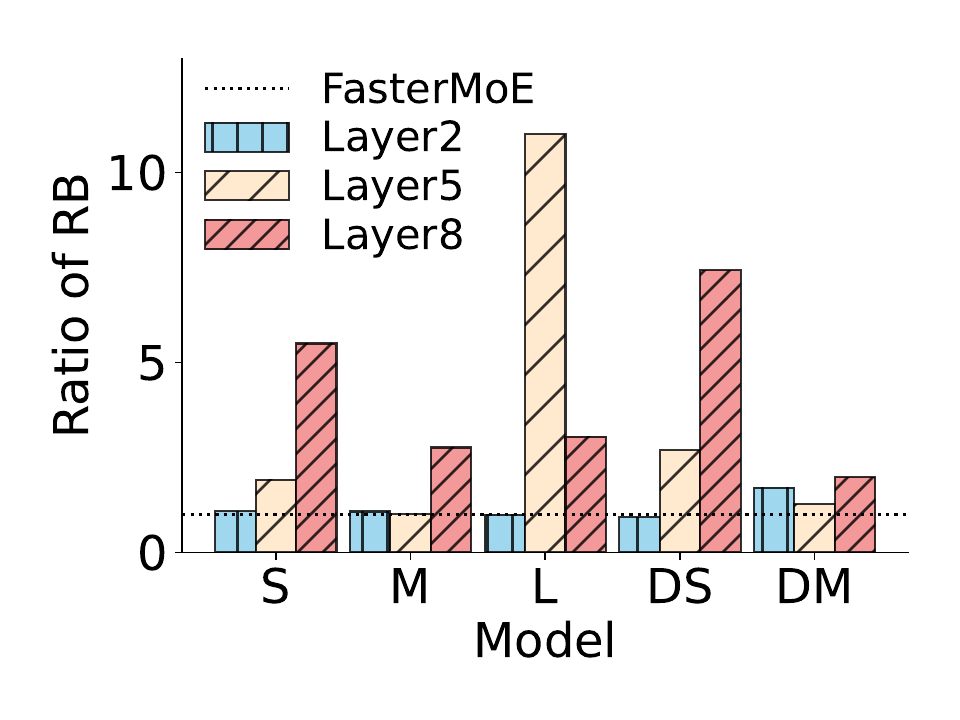}}
\quad  
\subfloat[\label={$k$=2}]{
\includegraphics[scale=0.25]{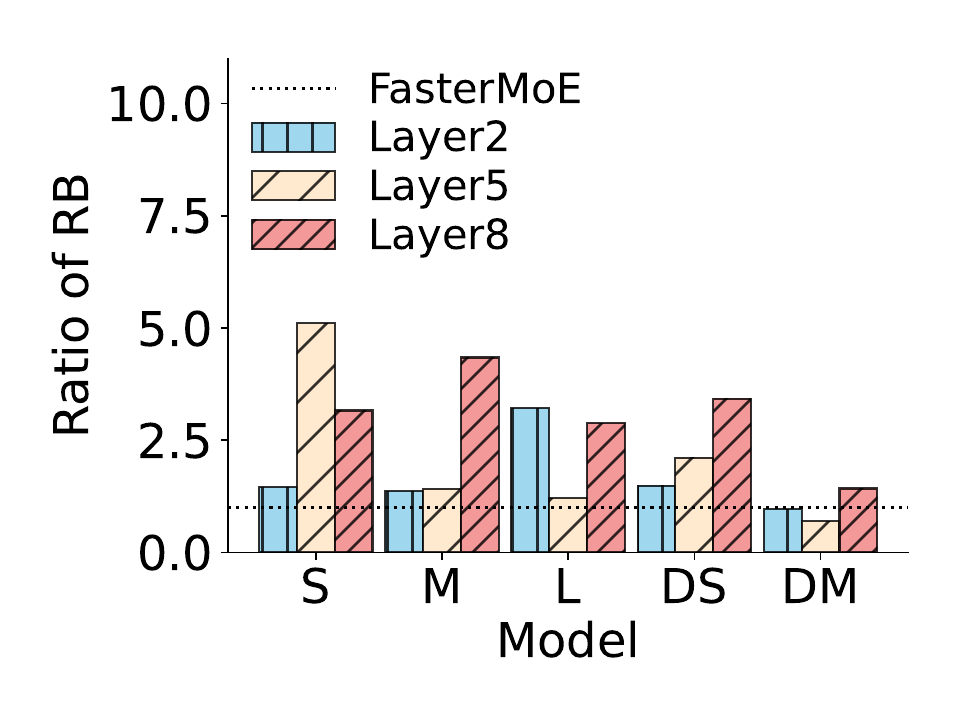}}
\caption{The ratio of \textit{RB} of the planner to FasterMoE in different models and $k$. The balance degree is the standard deviation of the input distribution tensor. The ratio of the balance degree before and after employing a load-balancing solution is \textit{RB} of the solution. The planner achieves up to 11.01x ratio of \textit{RB}.}
\label{fig:balance_degree}
\end{figure}

\textbf{Balance capability.} 
Balance capability serves as a key metric for load-balancing methods.
We define the balance degree as the standard deviation of the input distribution tensor. Besides, we denote the ratio of the balance degree before and after employing a load-balancing solution as \textit{RB} to describe its effect on the load. 

Fig.~\ref{fig:balance_degree} demonstrates the ratio of \textit{RB} of Pro-Prophet planner to that of fasterMoE on different layers in different values of $k$. It is worth mentioning that the training time for the planner is lower than that of FasterMoE. In most cases, the planner achieves a higher \textit{RB} than fasterMoE. A ratio of \textit{RB} up to 11.01x indicates its ability to enhance training efficiency by fully exploiting the potential of load balancing. 
In experimental conditions including $k$=1 with layer=2, $k$=2 with layers=2 and 5, the ratios of \textit{RB} are below 1, suggesting that the planner tailors expert placement to the actual load, preventing the unnecessary allocation of experts.

In summary, the planner can dynamically determine the load-balancing strategy to maximize training efficiency, showing its superior balance capability.




\section{Related Work}


\textbf{Hybrid parallelism.}
Hybrid parallelism strategies\cite{li2022amp, merak, hippie, li2023colossal, tarnawski2021piper, SSI-2023-0051} have been widely used to train large-scale dense models. These hybrid parallelism strategies consist of but are not limited to DP\cite{rajbhandari2021zero, ren2021zero,  
 zhao2023pytorch, zhang2022mics}, TP\cite{megatron, bian2021maximizing}, PP\cite{autopipe, hph}, and sequence parallelism (SP)\cite{li2021sequence, korthikanti2023reducing, jacobs2023deepspeed, liu2023ring}. Unfortunately, it is hard to efficiently train large MoE models utilizing these hybrid parallelism strategies.

To overcome the challenge, a series of works combine EP with above parallelism strategies and effectively improve training efficiency. 
Switch transformers combines EP with TP straightforwardly and designs a scheme to place the model and data on TPUs. 
Bagualu\cite{bagualu} develops a hybrid parallel strategy that integrates EP and DP tailored for high-performance computing architectures, along with communication and storage optimizations designed to enhance training efficiency. Deepspeed-MoE designs an effective combination of DP, EP and TP for inference but is easy to extend to model training. In the MoE layer, it introduces Allgather and Allreduce primitives to aggregate data and immediate results. 
Tutel also proposes DP, TP, and EP hybrid parallelism strategies and designs an adaptive parallelism switching method that enables O(1) overhead in runtime switching. Based on the Deepspeed-MoE and Tutel, Parm\cite{pan2024parm} combines DP, EP, and expert-slice parallelism (ESP) and proposes a fine-grained communication scheduling to improve the utilization of communication links. 
DeepSpeed-TED\cite{singh2023hybrid} designs a 3-dimensional hybrid parallelism strategy that contains DP of Zero-3, TP of Megatron-LM, and EP of Deepspeed-MoE. Besides, it proposes a memory and communication optimization for better scalability. 
The methods of Pro-Prophet are compatible with these hybrid parallelism strategies and can help further improve the training efficiency.


\textbf{Communication schedule.}
Overlapping communications and computing can enhance hardware utilization and improve the system's throughput~\cite{wfbp, pipetransformer}. 
Previous communication scheduling methods~\cite{merak, trirace} for dense models have demonstrated promising results. In this paragraph, We focus on introducing works designed for MoE models. 

Mainstream communication scheduling works focused on pipelining A2A and expert computation. Specifically, they partition an A2A and expert computation operation into sub-operators and overlap communication sub-operators with computation ones. 
Methods implemented on Gshard-like frameworks such as Lina\cite{li2023accelerating}, Tutel, ScheMoE\cite{shi2024schemoe}, and PipeMoE\cite{shi2023pipemoe} partition computation and communication operators based on the shape of expert computation matrix. FasterMoE is implemented on FastMoE and partitions operators into irregular sub-operators to schedule.
Pro-Prophet is compatible with these works as Pro-Prophet allows for overlapping communications and computations at the level of MoE blocks.


\section{Conclusion}
In this paper, we propose Pro-Prophet, a systematic load-balancing approach for efficient training of MoE models. 
We observe a locality among input distributions and use it to design the planner and scheduler. 
Pro-Prophet planner identifies lightweight expert placements and designs a locality-based greedy algorithm to efficiently search for a communication-efficient expert placement using its proposed performance model, effectively reducing the communication overhead.
Pro-Prophet scheduler predicts the input distribution based on the locality in the MoE model training and applies block-wise scheduling to overlap communications and computations, further decreasing the communication cost. 
Our experiments show that Pro-Prophet achieves 1.18-2.66x and 1.01-1.50x speedups compared to Deepspeed-MoE and FasterMoE.
Besides, Pro-Prophet achieves a load balancing enhancement of up to 11.01 when compared to FasterMoE.

\bibliographystyle{IEEEtran}
\bibliography{ref}

\vfill

\end{document}